\pdfoutput=1
\documentclass[fleqn,usenatbib,usedcolumn]{mnras}

\usepackage{graphicx,amssymb,hyperref}
\usepackage{color}

\usepackage[fleqn]{amsmath}
\usepackage{abbrev}
\usepackage{siunitx}
\usepackage{gensymb}
\usepackage{booktabs,tabularx}
\usepackage{flushend}
\usepackage{xspace}
\usepackage{xcolor}

\usepackage[shortlabels]{enumitem}
\usepackage{subcaption}
\setlist[itemize]{leftmargin=5pt}

\def\mnras{MNRAS}
\def\apj{ApJ}

\def\simlt{\lower.5ex\hbox{$\; \buildrel < \over \sim \;$}}
\def\simgt{\lower.5ex\hbox{$\; \buildrel > \over \sim \;$}}
\def\etal{{\it et al.}}

\def\kpc{\mathrm{\, kpc}}

\def\mpc{\mathrm{\, Mpc}}
\def\msun{\mathrm{\, M_\odot}}

\def\cmsg{\, \mathrm{cm^2 \, g^{-1}}}

\defcitealias{2012ApJ...761....1W}{W12}
\defcitealias{2019MNRAS.488.3646R}{R19}
\defcitealias{2021MNRAS.500.2627H}{H21}
\newcommand{\har}{\citetalias{2021MNRAS.500.2627H}\xspace}

\newcommand{\bahamas}{\textsc{bahamas}\xspace}

\def\gs{\mathrel{\raise1.16pt\hbox{$>$}\kern-7.0pt \lower3.06pt\hbox{{$\scriptstyle \sim$}}}}         
\def\ls{\mathrel{\raise1.16pt\hbox{$<$}\kern-7.0pt \lower3.06pt\hbox{{$\scriptstyle \sim$}}}}   

\newcommand{\vect}[1]{\boldsymbol{#1}}

\newcommand{\be}{\begin{equation}}
\newcommand{\ee}{\end{equation}}
\newcommand{\ba}{\begin{eqnarray}}
\newcommand{\ea}{\end{eqnarray}}

\DeclareGraphicsExtensions{.pdf,.png}

\title[Weak lensing shapes to test SIDM]{Why weak lensing cluster shapes are insensitive to self-interacting dark matter}

\author[A.\ Robertson \etal]{Andrew Robertson\thanks{e-mail: {\tt andrew.a.robertson@jpl.nasa.gov}}, Eric Huff and Katarina Markovi\v{c} 
\\Jet Propulsion Laboratory, California Institute of Technology, 4800 Oak Grove Drive, Pasadena, CA 91109, USA\\
}

\begin{document}

\maketitle

\label{firstpage}

\begin{abstract}

We investigate whether the shapes of galaxy clusters inferred from weak gravitational lensing can be used as a test of the nature of dark matter. We analyse mock weak lensing data, with gravitational lenses extracted from cosmological simulations run with two different dark matter models (CDM and SIDM). We fit elliptical NFW profiles to the shear fields of the simulated clusters. Despite large differences in the distribution of 3D shapes between CDM and SIDM, we find that the distributions of weak-lensing-inferred cluster shapes are almost indistinguishable. We trace this information loss to two causes. Firstly, weak lensing measures the shape of the projected mass distribution, not the underlying 3D shape, and projection effects wash out some of the difference. Secondly, weak lensing is most sensitive to the projected shape of clusters, on a scale approaching the virial radius ($\sim 1.5 \mpc$), whereas SIDM shapes differ most from CDM in the inner halo. We introduce a model for the mass distribution of galaxy clusters where the ellipticity of the mass distribution can vary with distance to the centre of the cluster. While this mass distribution does not enable weak lensing data to distinguish between CDM and SIDM with cluster shapes (the ellipticity at small radii is poorly constrained by weak lensing), it could be useful when modelling combined strong and weak gravitational lensing of clusters.

\end{abstract}


\section{Introduction}

Self-interacting dark matter (SIDM) is an alternative to the collisionless cold dark matter (CDM) that constitutes roughly 85\% of the mass in the standard cosmological model, $\Lambda$CDM. SIDM differs from CDM in that dark matter (DM) particles can interact with one another through some force in addition to gravity, at rates that are astrophysically interesting. These interactions are usually assumed to be elastic scattering, and in this context SIDM leads to a transport of heat towards the otherwise-cold inner regions of DM haloes \citep{2014PhRvL.113b1302K}, which typically lowers the DM density at the centre of haloes \citep[e.g.][]{2013MNRAS.430...81R}. This change in central density is a key prediction of SIDM, and there have been numerous attempts to measure and/or constrain the SIDM cross-section from observables related to the spherically averaged DM density profiles of haloes \citep[with][as some fairly recent examples]{2016PhRvL.116d1302K, 2019PhRvX...9c1020R, 2021JCAP...01..024S, 2021JCAP...01..043B, 2022MNRAS.510...54A, 2022arXiv220501123E, 2022arXiv220703506T}. The same interactions that transport heat through DM haloes, would also isotropize the orbits of DM particles, leading to DM haloes becoming rounder. Inferring how round galaxy clusters are using weak gravitational lensing, with the hope of using this to learn something about the nature of DM, is the subject of this paper.

\citet{2002ApJ...564...60M} was the first paper to present a constraint on the SIDM cross-section derived from the \emph{shapes} of DM haloes, where by `shapes' we mean how the haloes depart from being spherically symmetric, often defined in terms of the ellipticity (or equivalently by the axis ratio) of iso-density contours. Specifically, gravitational lensing by the galaxy cluster MS2137-23 suggested that the DM distribution at a radius of $70 \kpc$ was elliptical (i.e. it was not round). If one assumes that the halo should be round within the radius at which each particle has scattered once on average, then this leads to a constraint of $\sigma/m < 0.02 \cmsg$, which is an order of magnitude tighter than the tightest constraints derived from spherically averaged density profiles.

\citet{2013MNRAS.430..105P} used cosmological simulations to show that the \citet{2002ApJ...564...60M} constraint was overstated by over an order of magnitude. There are a number of reasons for this, including the fact that it takes more than one scattering per particle to erase the triaxial shape of a halo \citep{2017JCAP...05..022A}, and that gravitational lensing is sensitive to the projected mass distribution of the lens, which -- at some projected radius $R$ -- receives a contribution from the mass at all 3D radii, $r \geq R$. While finding that this previous lensing result was overstated, \citet{2013MNRAS.430..105P} did find that X-ray isophotes of NGC 720, a massive elliptical galaxy, were likely incompatible with a cross-section as large as $1 \cmsg$. Further X-ray observations of another 11 elliptical galaxies \citep{2021JCAP...05..020M} broadly supported the \citet{2013MNRAS.430..105P} findings, although further simulation work suggested that the inclusion of baryons into CDM and SIDM simulations makes the resulting shape distributions more similar to one another \citep{2022MNRAS.516.4543D}, such that a cross-section of $1 \cmsg$ is still consistent with the X-ray shapes of elliptical galaxies.

Returning to galaxy cluster scales, \citet{2018MNRAS.474..746B} ran DM-only zoom-in simulations of 28 galaxy clusters, each simulated with CDM as well as a couple of SIDM models. Focusing on their larger cross-section simulations (with $\sigma/m = 1 \cmsg$) they found that halo shapes differ from CDM at the $1 \sigma$ level (where $\sigma$ here refers to the intrinsic scatter of shapes for haloes simulated with a single DM model) out to 30\% of the virial radius, whereas density profiles differ by $1 \sigma$ only out to 5\% of the virial radius. \citet{2019MNRAS.488.3646R} also simulated galaxy clusters with different SIDM cross-sections, but this time included baryons and associated models for the physics of galaxy formation. They also found that shape differences in the DM distributions persisted out to fairly large radii. In addition they found that the shape differences between CDM and SIDM (with a cross-section of $1 \cmsg$) were significantly larger than the shape differences between DM-only and hydrodynamical simulations with the same DM model. This suggests that galaxy cluster shapes are relatively robust to the inclusion of baryonic physics. However, they found that changing the SIDM cross-section within the range currently allowed on cluster scales did not appreciably alter the shapes of the gas or galaxy distributions within clusters, suggesting that methods which are sensitive to the DM mass distribution (such as gravitational lensing) are necessary in order to use cluster shapes as a probe of SIDM.

As for observations of galaxy cluster shapes, \citet{2010MNRAS.405.2215O} measured the ellipticity of individual clusters, fitting elliptical NFW profiles to a sample of 18 X-ray luminous clusters and finding them to have a mean axis ratio of 0.54. This was in good agreement with the expectation at the time, from projecting the 3D axis ratio distributions measured in $\Lambda$CDM  $N$-body simulations \citep{2002ApJ...574..538J}. More recently, \cite{2018MNRAS.475.2421S} measured the ellipticity of SDSS clusters identified by the redMaPPer algorithm, finding a mean axis ratio of 0.56. This result relied on stacking galaxy cluster shear fields, and so was dependent on the assumption that the major axis of each cluster's mass distribution is aligned with the major axis of its galaxy distribution (which was used to orientate each cluster in the stack). Given the estimated uncertainties, this result is in agreement with the result for individual clusters, and also with more recent simulation results \citep[e.g.][]{2017MNRAS.466..181D} which had explicitly calculated projected shapes. The fact that observations of cluster shapes have, in general, been found to be in agreement with the $\Lambda$CDM expectation, could potentially rule out some SIDM models where haloes are expected to be rounder. However, while the shapes of simulated clusters with CDM and SIDM have been compared in 3D, there has been little work on the observational consequences of this in terms of weak lensing inferred shapes. This is what we investigate in this paper.






This paper is structured as follows. In Section~\ref{sect:sim} we briefly describe the simulations that we use throughout the rest of the paper. Then in Section~\ref{sect:3D_shapes} we present the algorithms that we use to measure the shapes of our simulated clusters, and discuss differences in the 3D shapes of CDM and SIDM clusters, before turning our attention to 2D (projected) shapes in Section~\ref{sect:2D_shapes}. In Section~\ref{sect:WL} we describe how to generate weak lensing maps for our simulated clusters, and then fit these with elliptical NFW profiles. We present the distribution of weak lensing-inferred axis ratios in CDM and SIDM, and discuss the radial scale on which weak lensing is most sensitive to the cluster shape. We also introduce a new mass distribution -- an NFW profile with a radially-varying ellipticity -- and show the results of fitting this to our sample of haloes. Finally, we conclude in Section~\ref{sect:conclusions}.

\section{The BAHAMAS-SIDM simulations}
\label{sect:sim}

To see how the expected weak lensing signal depends on the nature of DM, we use galaxy clusters extracted from the \bahamas-SIDM hydrodynamical simulations \citep{2019MNRAS.488.3646R}. These simulations used the same initial conditions and model of galaxy formation as one of the original (CDM) \bahamas simulations \citep{2017MNRAS.465.2936M,McCarthy2018}, but included SIDM interactions, with their implementation described in \citet{2017MNRAS.465..569R}.  \bahamas was run using a modified version of the {\sc Gadget-3} code \citep{Springel2005}.  The simulations include subgrid treatments for metal-dependent radiative cooling \citep{Wiersma2009a}, star formation \citep{Schaye2008}, stellar evolution and chemodynamics \citep{Wiersma2009b}, and stellar and AGN feedback \citep{DallaVecchia2008,Booth2009}, developed as part of the OWLS project (see \citealt{Schaye2010} and references therein). The simulations we use are of a periodic box, $400 \, h^{-1} \, \mpc$ on a side, with $2 \times 1024^3$ particles. They employ a WMAP 9-yr cosmology\footnote{With $\Omega_\mathrm{m}=0.2793$, $\Omega_\mathrm{b}=0.0463$, $\Omega_\mathrm{\Lambda}=0.7207$, $\sigma_8 = 0.812$, $n_\mathrm{s} = 0.972$ and $h = 0.700$.} \citep{2013ApJS..208...19H}, and have DM and (initial) baryon particle masses of $\num{5.5e9} \msun$ and $\num{1.1e9} \msun$, respectively. The Plummer-equivalent gravitational softening length is $5.7 \kpc$ in physical coordinates below $z=3$ and is fixed in comoving coordinates at higher redshifts.

For simplicity, we focus in this paper on comparing just two DM models: the largest SIDM cross-section used in \bahamas ($1 \cmsg$, which we label SIDM1), and CDM. For each of these models we extract the 100 most massive \emph{friends-of-friends} groups\footnote{For a description of the friends-of-friends algorithm, see e.g. \citet{2011ApJS..195....4M}} from the $z=0.375$ snapshot. These have halo masses that approximately span the range $14.3 < \log_{10} M_{200} / \msun < 15.3$, where $M_{200}$ is the mass enclosed within a spherical aperture with radius $r_{200}$, and $r_{200}$ is the radius within which the mean enclosed density is equal to 200 times the critical density. Note that previous work has shown that average halo shapes are somewhat mass dependent. For example, in \citet{2007MNRAS.376..215B} the median minor-to-major axis ratio decreases by about 0.05 when moving from halo masses of $10^{13.5} \msun$ to $10^{14.5} \msun$. This amount is small compared with the width of the axis ratio distributions we find in this work, indicating that our distributions primarily reflect scatter at fixed mass, as opposed to a range of halo masses leading to a range of axis ratios.

\section{3D shapes as a function of radius}
\label{sect:3D_shapes}

Before discussing the shapes of clusters as inferred from weak lensing, we first discuss the ``true'' shapes of clusters calculated directly from the 3D positions of particles in our simulated haloes. We calculate the shapes of the DM, stellar and gas distributions, including all particles of the relevant type in each calculation. This means that, for example, the stellar distribution of a cluster includes the stars of the central galaxy and any other cluster member galaxies, as well as what might be considered the intracluster light.

\subsection{Inertia tensor shapes}
\label{sect:inertia_tensor_3D}

Following previous work on the shapes of simulated dark matter haloes \citep[e.g.][]{2006MNRAS.367.1781A, 2012JCAP...05..030S, 2017MNRAS.466..181D}, we calculate the shapes of the different components (DM, stars and gas) of our simulated haloes using algorithms that make use of the \emph{inertia tensor}
\begin{equation}
I_{ij} \equiv \sum_n x_{i,n} \, x_{j,n} \, m_{n} \, \Big/  \, \sum_n m_{n}
\label{mass_tensor}
\end{equation}
where $(x_{1,n},x_{2,n},x_{3,n})$ are the coordinates of the $n$th particle, which has mass $m_{n}$. We define positions of particles with respect to the centre of the halo, for which we use the location of the most gravitationally bound particle calculated by the SUBFIND algorithm \citep{2001MNRAS.328..726S}. All spherical/ellipsoidal search volumes are centred on this most bound particle also. 

\subsubsection{Standard inertia tensor shape}

To calculate what we call the \emph{standard} inertia tensor shape at a certain radius, $r_\mathrm{shape}$, we begin by finding all particles in a sphere of radius $r_\mathrm{shape}$, and then calculate $I_{ij}$ for this set of particles. We label the eigenvalues of $I_{ij}$ as $a^2$, $b^2$ and $c^2$, with $a \geq b \geq c$, such that $a$ is the standard deviation of particle positions along the major axis, while $c$ is the equivalent along the minor axis.

From the eigenvalues, an ellipsoid can be defined which has a minor-to-major axis ratio of $c/a$ and an intermediate-to-major axis ratio of $b/a$, where the principal axes of the ellipsoid align with the corresponding eigenvectors of $I_{ij}$. Our process is iterative, where in each step $I_{ij}$ is calculated for all particles within the ellipsoid found in the previous step. We keep the volume enclosed by the ellipsoid equal to the original volume of the sphere (i.e. $4\pi r_\mathrm{shape}^3 / 3$), and stop the iteration when subsequent iterations agree on both axis-ratios ($c/a$ and $b/a$) to better than 1\%. We note that this iterative procedure leads to the correct inference on the axis ratios of mass distributions with known axis ratios (for example, made by stretching a spherically symmetric particle distribution by a known amount along orthogonal axes). This is not the case if one simply calculates the inertia tensor for all particles within a spherical volume, which biases the axis ratios towards unity. This is demonstrated in Appendix~B of \citet{2019MNRAS.488.3646R}.

\subsubsection{Inertia tensor in shells}

An alternative definition of the shape of the halo at a given radius considers only the mass at (or close to) that radius, as opposed to all mass enclosed. The procedure to calculate such a shape is similar to the one described above, but the particles used to calculate $I_{ij}$ in a particular step are those in an ellipsoidal shell. These particles are defined as those with an ellipsoidal radius, $r_e$, that is between $f \times r_\mathrm{shape}$ and $r_\mathrm{shape} / f$, where $f$ is a number slightly smaller than unity, that controls the width of the shell. The ellipsoidal radius for a particle at position $\vect{x}$ is
\begin{equation}
r_e = \sqrt{(\vect{x} . \vect{e_a})^2/a^2 + (\vect{x} . \vect{e_b})^2/b^2 +(\vect{x} . \vect{e_c})^2/c^2} \, \left( a b c \right)^{1/3} ,
\label{elliptical_radius}
\end{equation}
with $\vect{e_a}$ the eigenvector associated with the largest eigenvalue, $a^2$, of $I_{ij}$ from the previous step of the iteration, and similar for $\vect{e_b}$ and $\vect{e_c}$.

\subsubsection{Reduced inertia tensor}
\label{sect:reduced_Iij}

Another alternative shape definition again considers all particles within some ellipsoid (as done for the standard inertia tensor shape), but replaces the inertia tensor with the \emph{reduced inertia tensor}
\begin{equation}
\tilde{I}_{ij} \equiv \sum_n \frac{x_{i,n} \, x_{j,n} \, m_{n}}{r_{e,n}^2} \, \Big/  \, \sum_n m_{n}
\label{reduced_inertia_tensor}
\end{equation}
where $r_{e,n}$ is the ellipsoidal radius of the $n$th particle, defined in Equation~\eqref{elliptical_radius}. This procedure gives more weight to particles closer to the centre than the standard inertia tensor does.

\subsection{Comparison of shape definitions}

\begin{figure}
        \centering
        \includegraphics[width=\columnwidth]{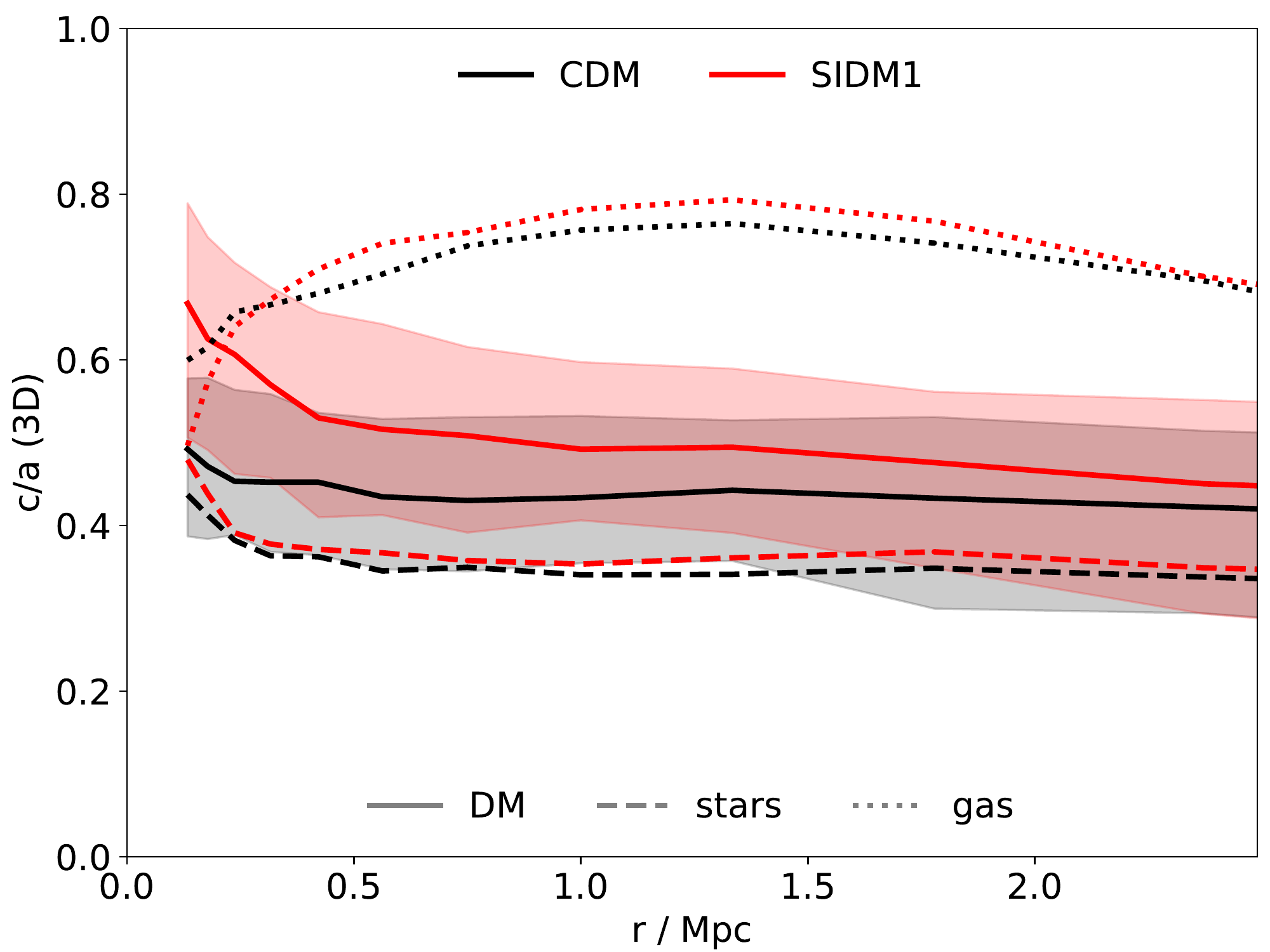}     
	\caption{Median 3D minor-to-major axis ratios as a function of radius for the 100 most massive galaxy clusters in the $z=0.375$ snapshots from \bahamas-CDM and \bahamas-SIDM1. The axis ratios were calculated using the reduced inertia tensor (Section~\ref{sect:reduced_Iij}). The shaded regions denote the 16th to 84th percentile spread of the DM axis ratios.}
	\label{fig:3D_axis_ratios}
\end{figure}

It is not obvious which of these three algorithms to calculate axis ratios as a function of radius (using either the standard inertia tensor, the inertia tensor in shells, or the reduced inertia tensor) should be preferred, and all three have been used previously in the literature. Note that different shape definitions will typically lead to different axis ratios, and can paint slightly different pictures about how distinguishable different DM models are. For example, the difference in median DM $c/a$ between CDM and SIDM1 (shown in Fig.~\ref{fig:3D_axis_ratios}) is 0.037 at $r = 2 \mpc$ when the reduced inertia tensor is used, whereas it is only 0.017 when calculated in shells, because in the former case the shape at large $r$ depends also on the mass distribution at small radii -- where CDM and SIDM1 are most different. Ultimately, we are interested in the shapes of haloes as they can be measured from observations, and it turns out that the reduced inertia shapes are the most correlated with the weak lensing shapes that we calculate later in the paper (Section~\ref{sect:scale_of_WL}). For this reason we use reduced inertia tensor shapes unless otherwise specified.



\subsection{Joint distribution of DM and stellar shapes}

\begin{figure*}
  \centering
\begin{subfigure}[b]{0.5\linewidth}
    \centering
\includegraphics[width=\textwidth]{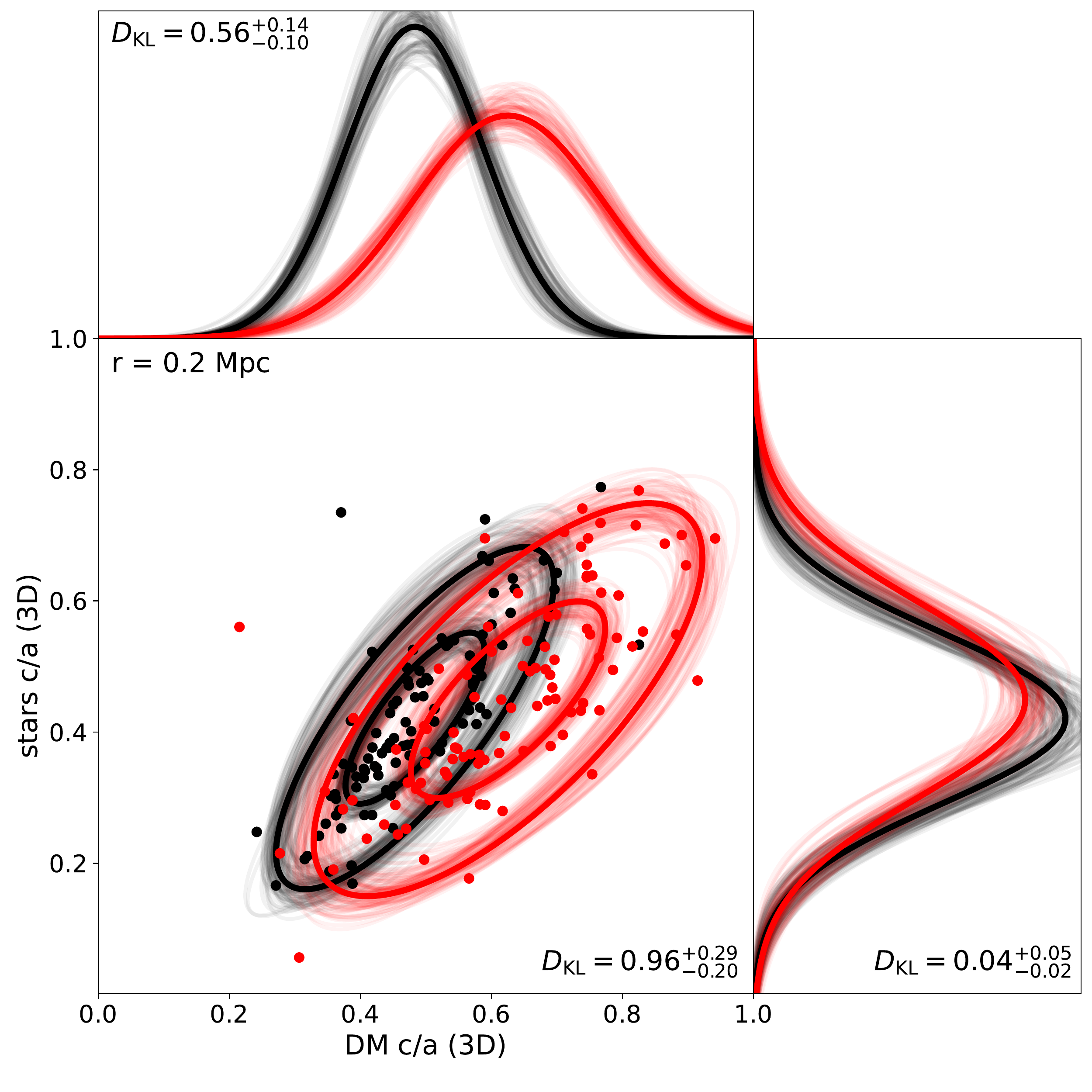}
\end{subfigure}%
\begin{subfigure}[b]{0.5\linewidth}
    \centering
\includegraphics[width=\textwidth]{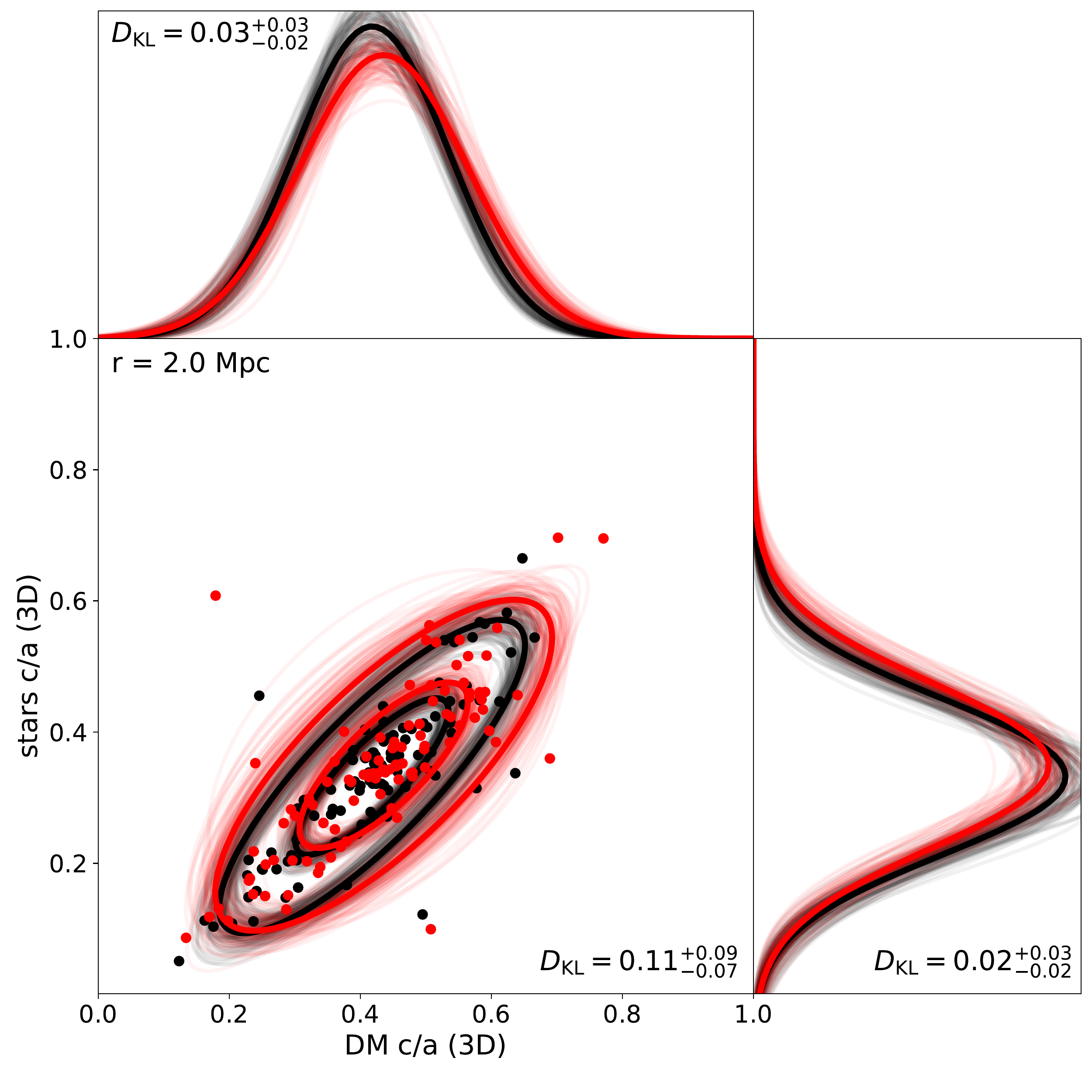}
  \end{subfigure} 
  \caption{Left: the joint distribution of DM and stellar minor-to-major axis ratios measured in 3D using the reduced inertia tensor at a radius of 200 kpc. Black and red points are for the 100 most massive friends-of-friends groups from each of the CDM and SIDM1 \bahamas simulations, respectively. The ellipses show the 1 and 2$\sigma$ contours of the multivariate Gaussian\protect\footnotemark \, with the same mean and covariance matrix as the 100 points, while the panels above and to the right show the 1D projections of these. Each panel contains the value of KL divergence associated with the difference between the respective CDM and SIDM1 distributions. The faded ellipses and 1D Gaussians are the same as the solid lines, but for many bootstrap samples of the 100 haloes, giving an indication of the uncertainty on these multivariate Gaussians due to the finite number of simulated haloes. Right: the same as the left panel but with shapes measured at a radius of 2 Mpc where the CDM and SIDM1 distributions now look much more similar to one another.}
  \label{fig:3D_joint_dist}
\end{figure*}

To give a better indication of what the curves in Fig.~\ref{fig:3D_axis_ratios} actually mean for distinguishing between the two DM models (CDM and SIDM1) using cluster shapes, we plot the distribution of shapes at fixed radii in Fig.~\ref{fig:3D_joint_dist}. We plot the joint distributions of stellar and DM $c/a$ at two radii, 200 kpc and 2 Mpc. The left panel of Fig.~\ref{fig:3D_joint_dist} (with shapes at 200 kpc), demonstrates an interesting phenomenon, whereby even though the distribution of stellar shapes is very similar in the two different DM models, adding information about the stellar shapes to the DM shapes increases the ease of discriminating between CDM and SIDM1 considerably, when compared with using DM shapes alone.
This can be quantified by the Kullback–Leibler (KL) divergence \citep{Kullback:1951zyt}, which for two probability distributions, $p(x)$ and $q(x)$, is defined as 
\begin{equation}
D_\mathrm{KL} = \int p(x) \log \left( \frac{p(x)}{q(x)} \right) \mathrm{d}x.
\label{eq:KL}
\end{equation}

The KL divergence is the expected value of the difference between the logarithms of the probability densities $p$ and $q$, where the expectation is taken over the probability distribution $p$. For example, if one were to make $N$ random draws from the probability distribution $p(x)$ (which we label the data, $D$) and then calculate the likelihood for two models -- $M_1$: the samples were drawn from $p(x)$ and $M_2$: the samples were drawn from $q(x)$ -- the expected value of the $\log$ of the Bayes factor, $\left< \log \mathcal{B} \right> = \left< \log P(D | M_1) / P(D | M_2) \right> = N D_\mathrm{KL}$. This makes $D_\mathrm{KL}$ a useful metric for determining how dissimilar -- and therefore how easy to distinguish -- the CDM and SIDM1 distributions of axis ratios are. It can also be calculated for multi-dimensional random variables (for example, the combination of the DM $c/a$ and the stellar $c/a$, or the DM $c/a$ at multiple radii), and meaningfully be compared between situations that have different numbers of dimensions.

Given our fairly limited number of high-mass clusters, we do not try to estimate the probability distribution of shapes directly,\footnote{For example, using kernel density estimation.} but instead assume that the distribution of shapes can be adequately described by a multivariate Gaussian. We therefore calculate the mean and covariance matrix of the axis ratio distributions within each DM model, which are indicated by the elliptical contours in Fig.~\ref{fig:3D_joint_dist}. The marginalised distribution of DM and stellar axis ratios are also plotted in Fig.~\ref{fig:3D_joint_dist} following this Gaussian approximation. A nice feature of multi-variate Gaussians is that the KL divergence between two of them can be calculated analytically \citep[e.g.][]{Duchi2016DerivationsFL}, which is how we calculate the $D_\mathrm{KL}$ values shown on the plot. Note that in addition to calculating the mean and covariance matrix for our full samples of 100 haloes, we also generate bootstrap samples (i.e. samples of 100 haloes, drawn from the full 100 with replacement) and calculate their means and covariances. The corresponding elliptical contours and marginalised 1D Gaussian distributions are plotted as faded lines in Fig.~\ref{fig:3D_joint_dist}, which give an indication of the uncertainty on the distributions with each DM model due to the sample size of 100 haloes. The errors quoted on the $D_\mathrm{KL}$ values cover the 16th - 84th percentiles of the $D_\mathrm{KL}$ values for the bootstrapped samples. We emphasize that the results of our bootstrapping procedure are not used in any future analysis, but instead are to guide the eye as to the significance of the CDM vs SIDM1 shape distribution differences given the relatively small sample of simulated haloes.

Taking the example of the shapes of the DM and stellar distributions at 200 kpc, we can see that the distribution of DM shapes is fairly different between the two models ($D^\mathrm{DM}_\mathrm{KL} = 0.56$), while the distribution of stellar shapes is much more similar ($D^\mathrm{star}_\mathrm{KL} = 0.04$). Nevertheless, adding the information about the stellar shapes to the DM shapes increases the KL divergence fairly considerably over the case of only the DM shapes ($D^\mathrm{DM,star}_\mathrm{KL} = 0.96$). So if one were hoping to find a log Bayes factor of greater than 10 \citep[usually considered ``very strong'' evidence, e.g.][]{2013JCAP...08..036N} in favour of CDM over SIDM1 (and one lived in a CDM universe) then we would expect to need around 18 systems with accurately measured DM shapes, or 250 systems with accurately measured stellar shapes, or just 11 systems with both DM and stellar shapes.

Note that in the discussion so far about the number of systems it would take to confidently distinguish between a CDM and SIDM1 universe we have assumed that DM and stellar shapes of systems can be accurately measured, and that the reason multiple systems are required is that the distribution of axis ratios for any particular DM model is broad, with significant overlap between the CDM and SIDM1 distributions. In practice, observations will only be able to measure these shapes to some level of statistical accuracy (see Section~\ref{sect:eNFW_fit}), and so this assumption is a very optimistic best case scenario. Also, these simulations are only an approximate prediction of what the shape distributions would look like with different DM models, and (for example) different implementations of baryonic physics could quite possibly lead to changes that are larger than the differences seen between the stellar $c/a$ distributions for CDM and SIDM1 at 200 kpc (so that in practice no number of observed systems would allow us to confidently determine the nature of DM from the shape of the stellar distribution alone). Despite these caveats, we continue to use the KL-divergence throughout this paper as a means of measuring how different the distributions of various axis ratios are between CDM and SIDM1.

\footnotetext{Here the $n \sigma$ contour refers to the contour where the probability density has dropped by a factor of $\exp -n^2/2$ with respect to the maximum probability density. In 2D, this means that the 1$\sigma$ contour does not contain approximately 68\% of the total probability, as it would in a 1D case.}

Turning to the case of shapes at larger radii, the right panel of Fig.~\ref{fig:3D_joint_dist} shows the joint DM-stellar shape distribution measured in 3D at a radius of 2 Mpc. As seen in Fig.~\ref{fig:3D_axis_ratios}, the DM shape distributions in CDM and SIDM1 become more similar when considering larger radii, but the challenge of using shapes at large radii to constrain the nature of DM is perhaps more obvious in Fig.~\ref{fig:3D_joint_dist}, where $D^\mathrm{DM}_\mathrm{KL}$ at 2 Mpc is 0.03 in contrast with 0.56 at 200 kpc, such that it is (in a sense) approximately twenty times more difficult to distinguish between CDM and SIDM1 using DM shapes measured on the scale of 2 Mpc compared with a scale of 200 kpc.

In addition to the DM and stellar shapes, Fig.~\ref{fig:3D_axis_ratios} also includes the gas shapes. In a similar manner to the stars, the gas shapes alone do not distinguish between DM models, but they do enhance the distinguishability of CDM and SIDM1 when combined with the DM shapes. However, the gas shapes are less correlated with the DM shapes than the stellar shapes are, such that the improvement is more modest.\footnote{In 3D at a radius of 200 kpc, $D^\mathrm{DM,gas}_\mathrm{KL} = 0.77$ in contrast with $D^\mathrm{DM,star}_\mathrm{KL} = 0.96$.} In addition, the combination of DM, star and gas shapes offers little improvement over just DM + star shapes. 

\section{2D shapes as a function of radius}
\label{sect:2D_shapes}

One major difference between the 3D shapes that can be measured in simulations, and the shapes of haloes that can be accessed observationally, is that observed data is almost always a 2D projection of some quantity. We therefore look at how the shapes of the projected DM, stellar and gas distributions differ between clusters simulated with CDM and SIDM1. The shapes were calculated using a 2D analogue of the reduced inertia tensor calculation described in Section~\ref{sect:reduced_Iij}. For the projected mass distributions, the 3D positions of all particles within a radius of $5 \times r_{200}$ of the most gravitationally bound particle in each cluster were projected along a line-of-sight.

\begin{figure}
        \centering
        \includegraphics[width=\columnwidth]{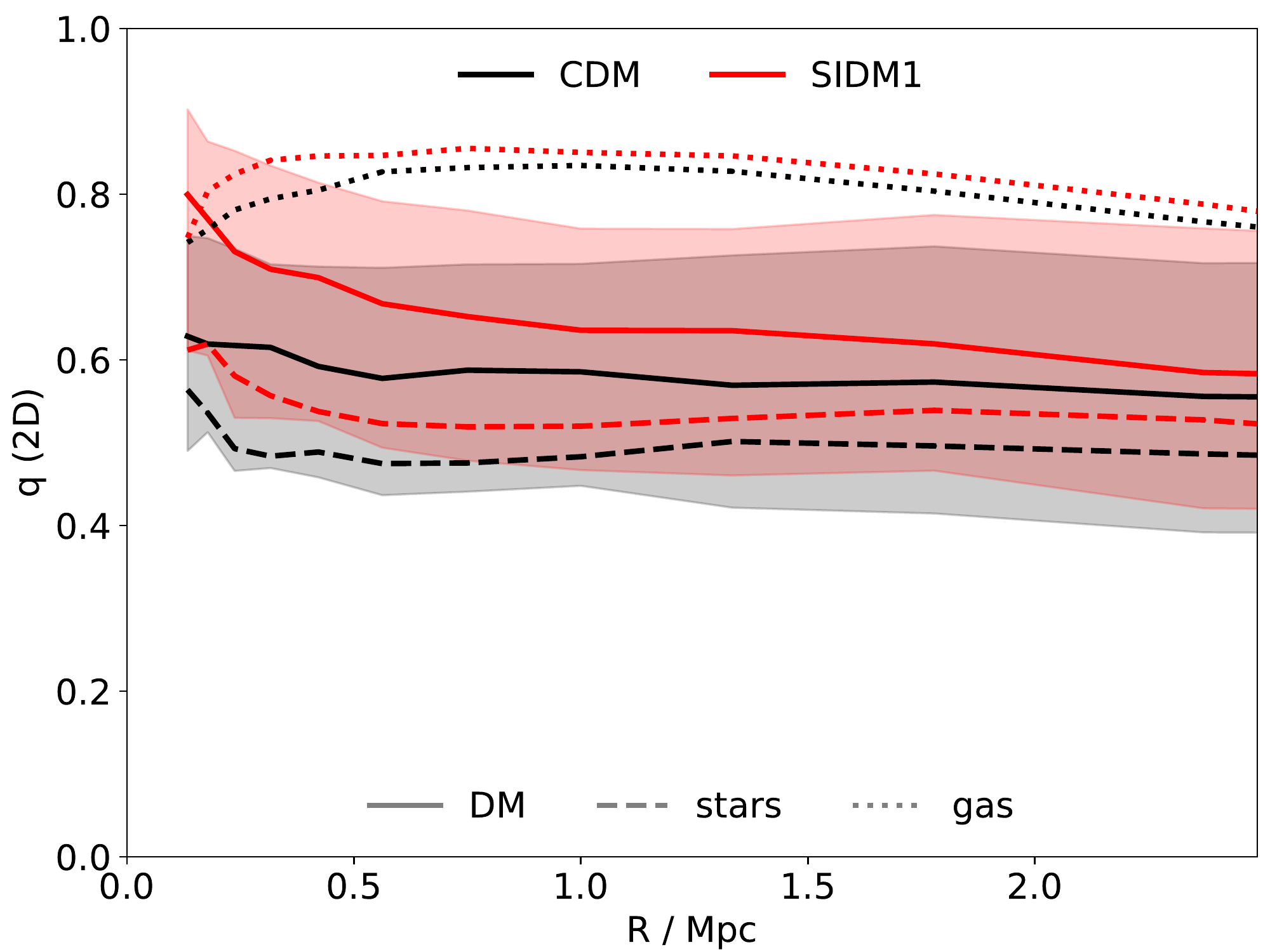}     
	\caption{Median 2D axis ratios as a function of projected radius for the 100 most massive galaxy clusters in the $z=0.375$ snapshots from \bahamas-CDM and \bahamas-SIDM1. The axis ratios were calculated using the 2D reduced inertia tensor. The shaded regions denote the 16th to 84th percentile spread of the DM axis ratios.}
	\label{fig:2D_axis_ratios}
\end{figure}

Fig.~\ref{fig:2D_axis_ratios} is the 2D analogue of Fig.~\ref{fig:3D_axis_ratios}, in which the simulated haloes have been projected along the simulation $z$-axis. Unsurprisingly, all cluster components are rounder in 2D than in 3D, because the 3D axis ratios were the ratios of minor and major axis lengths (i.e. were the most extreme axis ratio). Only in the case where the minor and major axes both lie in the plane of the sky would we expect the 2D axis ratio to match the 3D one, while in other cases the 2D shape will be rounder.

\begin{figure*}
  \centering
\begin{subfigure}[b]{0.5\linewidth}
    \centering
\includegraphics[width=\textwidth]{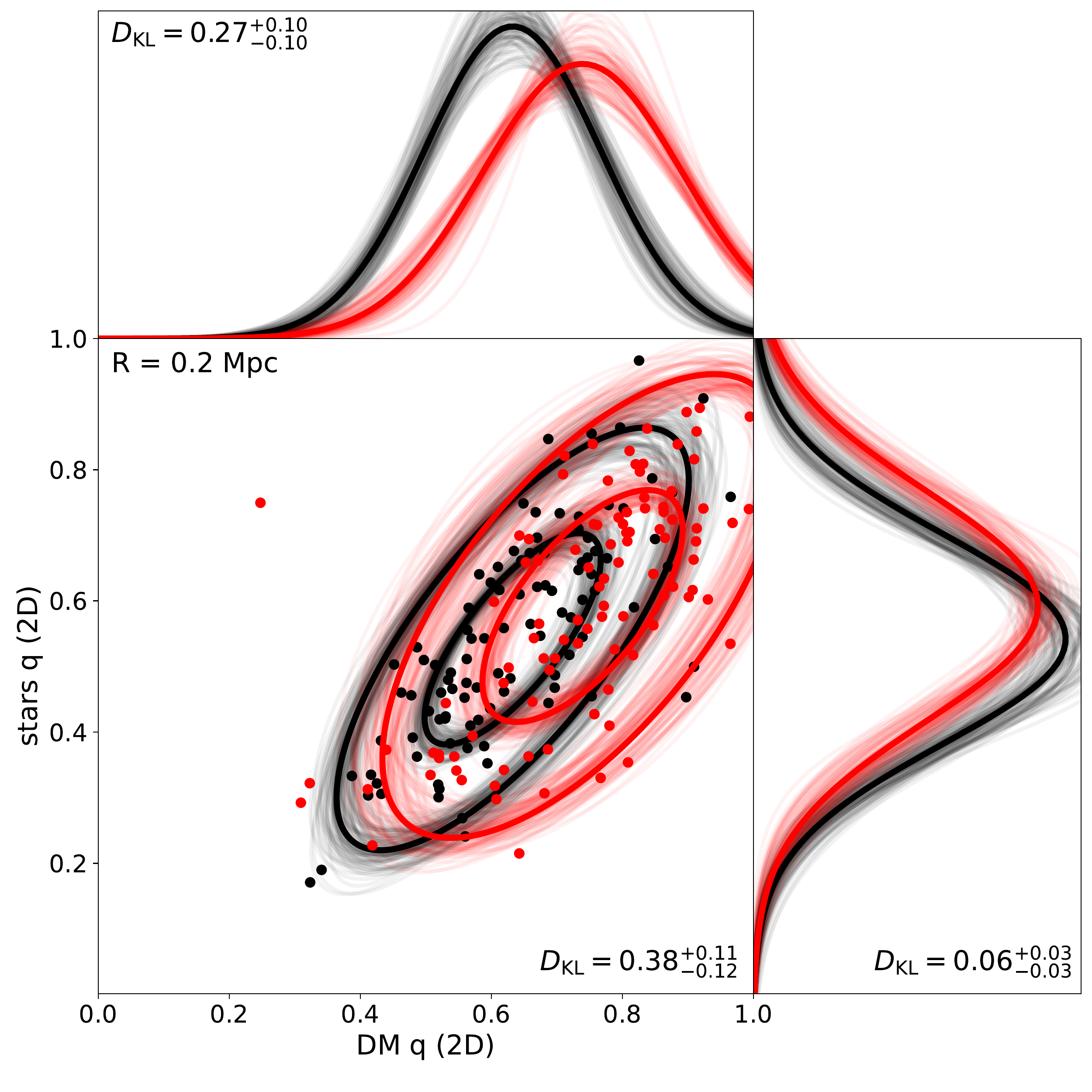}
\end{subfigure}%
\begin{subfigure}[b]{0.5\linewidth}
    \centering
\includegraphics[width=\textwidth]{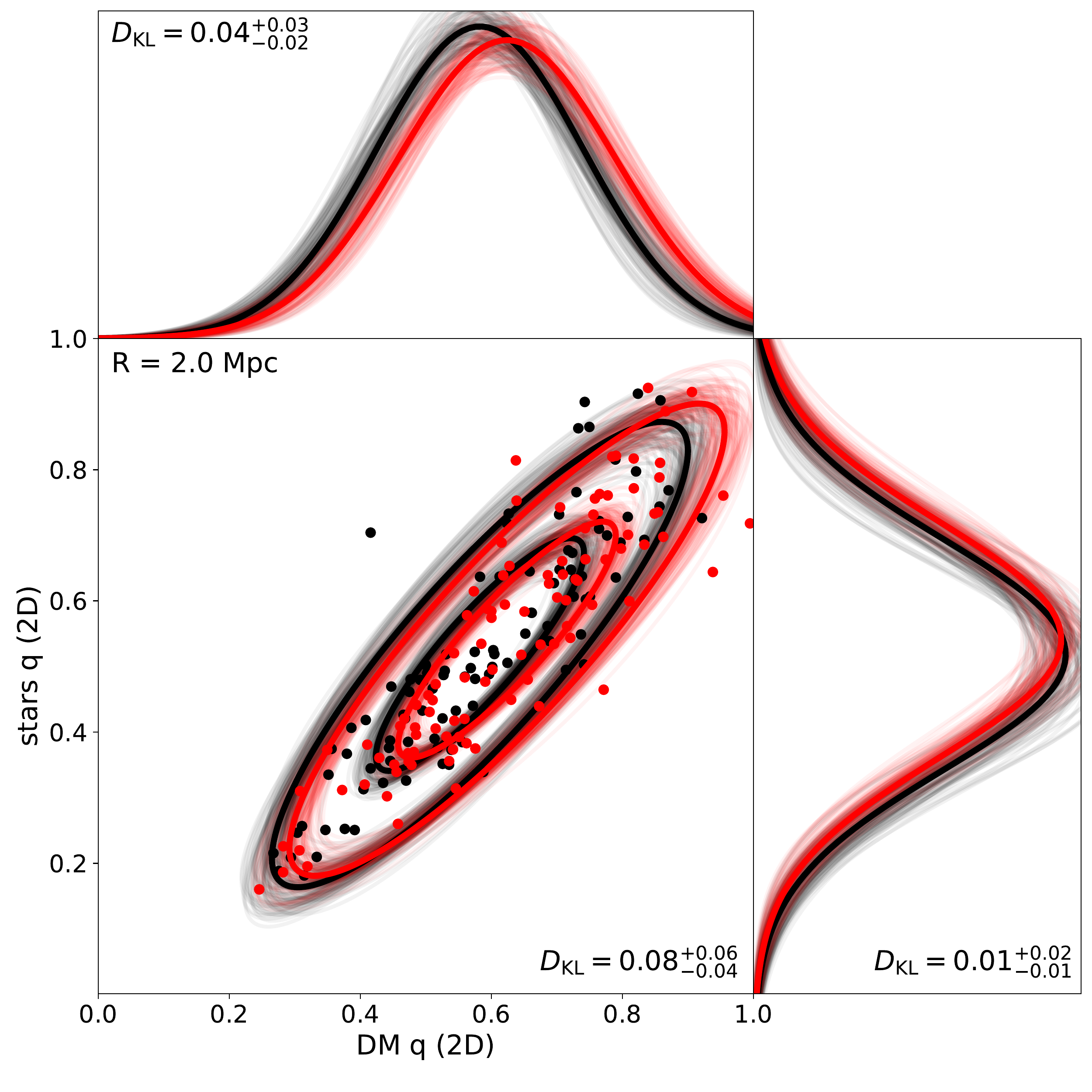}
  \end{subfigure} 
  \caption{The same as Fig.~\ref{fig:3D_joint_dist}, but now for 2D reduced inertia tensor axis ratios measured at a radius of $R = 200 \kpc$ (left) and $R = 2 \mpc$ (right). The 2D axis ratios are shifted towards being more round than their 3D counterparts, and the differences between CDM (black) and SIDM1 (red) are reduced by projection effects.}
  \label{fig:2D_joint_dist}
\end{figure*}

As well as both the CDM and SIDM1 axis ratios being rounder for 2D as opposed to 3D shapes, the distributions of DM shapes are now also harder to distinguish, with a drop of $D_\mathrm{KL}$ from 0.56 to 0.27 (at a fixed radius of $200 \kpc$), implying that roughly half of the constraining power from DM shapes is lost when going from 3D to 2D minor-to-major axis ratios. We discuss the cause of this next.

\subsection{The effects of projection}

There are two main effects that projection has on the shapes of haloes. The first, as described above, is that the 2D shape is some line-of-sight-dependent combination of the major, intermediate and minor axis lengths. The second is that projection mixes different radial scales, in that the projected mass at some 2D radius, $R$, receives a contribution from the 3D density at all 3D radii with $r > R$. Given that the 3D axis ratios become more similar at larger radii, this second effect will tend to decrease the distinguishability of SIDM1 projected shapes compared with those in CDM.

While it is not necessary to consider these processes separately, we think it is interesting to note that the first of these two effects can account for the bulk of the difference in $D^\mathrm{DM}_\mathrm{KL}$ between 3D shapes measured at $r = 200\,$kpc and 2D shapes measured at $R = 200\,$kpc. This claim is based upon predicting the distribution of 2D shapes by analytically projecting the 3D shapes (described by minor, major and intermediate axis lengths) along random lines-of-sight. At 200$\,$kpc, the KL divergence between 3D DM shapes is 0.56. When projecting the 3D distributions to calculate 2D axis ratios (with the mathematical details explained in Appendix~\ref{projecting_shapes}) this drops to 0.29, which is similar to the value of 0.27 for the true 2D DM shapes.

\section{Weak gravitational lensing shapes}
\label{sect:WL}

While the 2D shapes of clusters are more similar between CDM and SIDM1 than the 3D shapes, there are still differences in 2D shapes between these two DM models, at least at small to intermediate radii. In order to test whether these differences can be inferred from weak lensing data, we make mock weak lensing maps from our simulated clusters. We anchor our choices about the map size and number density of weakly lensed sources to the observations that will be made by SuperBIT, a balloon-borne visible-to-near-UV telescope, with an expected science flight in 2023 \citep{2018SPIE10702E..0RR}. In particular, we make maps of the weak lensing shear field with a $30 \times 30$ arcmin field-of-view,\footnote{Note that this is larger than the $15 \times 23$ arcmin field-of-view of SuperBIT camera's \citep{2022arXiv221009182S}, but would be achievable by tiling observations.} and assume that the dominant source of noise is \emph{shape noise} (the fact that galaxies without lensing have intrinsic shapes that are not round), which means that the noise level is set by the number density of source galaxies, for which we adopt 30 galaxies per square arcmin (McCleary et al. in prep.).


\subsection{Calculating gravitational shear maps}

Our method for calculating a mock shear field for a simulated cluster begins by generating a projected density map, $\Sigma$, for which we project the cluster's mass distribution along the simulation $z$-axis. To mitigate the effects of particle noise, we smooth out the mass from each particle in an adaptive manner, where the size of the smoothing kernel is smaller in higher density regions. We use an adaptive triangular shaped cloud scheme,\footnote{Implemented in the Python package \textsc{pmesh} \citep{yu_feng_2017_1051254}.} where each particle's mass is smoothed out in both the $x$ and $y$ directions by a triangular kernel with a full width of $2 \, r_{32}$, where $r_{32}$ is the 3D distance to a particle's 32nd nearest neighbour of the same particle species (so dark matter, gas and stars are each treated separately).

We then convert $\Sigma$ to the dimensionless convergence, $\kappa$, by dividing by the critical surface density for lensing $\Sigma_\mathrm{crit}$. This is defined as
\begin{equation}
\Sigma_\mathrm{crit} = \frac{c^2}{4 \pi G} \frac{D_s}{D_l D_{ls}},
\label{eq:sigma_crit}
\end{equation}
where $D_s$, $D_l$ and $D_{ls}$ are the angular diameter distances from the observer to the source, from the observer to the lens, and from the lens to the source, respectively. Note that for the lensing calculations in this paper we use the same  WMAP 9-yr cosmology \citep{2013ApJS..208...19H} as was used to run the simulations, with an assumed source redshift of $z_\mathrm{s} = 2$, and lens redshift $z_\mathrm{l} = 0.375$ (which is the redshift of the snapshots from which we extract the simulated clusters). This lens redshift is typical of the best observed galaxy clusters \citep[e.g.][]{2012ApJS..199...25P, 2017ApJ...837...97L}, while the source redshift is on the high-end of the redshift distribution expected for SuperBIT sources (McCleary et al. in prep.). Higher redshift sources lead to lower $\Sigma_\mathrm{crit}$ and subsequently more pronounced gravitational lensing, so this is a conservative choice with regards to our key finding that CDM and SIDM1 will be virtually impossible to distinguish from weak lensing inferred galaxy cluster shapes.  

To calculate the shear field, $\vect{\gamma}$, we use the fact that both $\kappa$ and $\vect{\gamma}$ depend on second derivatives of the projected gravitational potential. This means that the relationship between the Fourier transforms of $\kappa$ and $\vect{\gamma}$ is a simple one, and we calculate $\vect{\gamma}$ from $\kappa$ using discrete Fourier transforms \citep[for details see][]{2019MNRAS.488.3646R}. The initial resolution of our $\kappa$ maps is $512 \times 512$ pixels, leading to a pixel size of 3.5 arcsec (corresponding to $19 \kpc$ in the lens plane), which is also the resolution of our resulting $\vect{\gamma}$ maps. However, this resolution is considerably higher than the resolution at which shear can actually be measured -- 30 galaxies / arcmin$^2$ corresponds to one galaxy per (11 arcsec)$^2$. To save computation time, we therefore bin up our $\vect{\gamma}$ maps to a lower resolution of $240 \times 240$ pixels, corresponding to a pixel size of 7.5 arcsec (which is $39 \kpc$ at the lens redshift). To check that our findings are not sensitive to this pixel scale, we repeated our analysis at a lower resolution ($80 \times 80$ pixels), which produced very similar results.

\subsection{Fitting elliptical NFW profiles}
\label{sect:eNFW_fit}

In order to find a weak lensing-measured axis ratio for each simulated cluster, we fit an elliptical Navarro-Frenk-White (eNFW) profile to each cluster's mock shear field. Starting from the convergence due to a spherically symmetric NFW profile \citep{1997ApJ...490..493N}, $\kappa_\mathrm{NFW}(R)$, an eNFW profile replaces $R$ with $R_e$, a 2D elliptical radius defined by 
\begin{equation}
R_e = \sqrt{x^2 q + y^2/q},
\label{eq:2D_re}
\end{equation}
where $x$ and $y$ are distances along the major and minor axes respectively, and $q$ is the axis ratio. To calculate the associated shear field for an eNFW mass distribution we use \textsc{PyAutoLens} \citep{pyautolens}.

When generating these mock shear fields we ignore sources of noise, specifically, we do not add any shape noise.\hspace{-2pt}\footnote{Although we do use the expected level of shape noise for SuperBIT observations when defining our likelihood - which is then reflected in the width of the posteriors on our eNFW model parameters.} We also ignore sources of systematic error. For example, we generate and fit to maps of the shear, as opposed to calculating the reduced shear ($g = \gamma / (1 - \kappa)$) which is proportional to the change in the ellipticity of lensed galaxy images, and so more directly related to what is observable. This is because we wish to first demonstrate whether or not there is a signal to extract, before worrying about how one might extract it in the presence of noise and potential systematic errors.

In order to fit the eNFW model to our simulated shear fields, we need to define a likelihood function. We label the two components of shear $\gamma_1$ and $\gamma_2$, and then -- given our assumption that shape noise is the only source of noise -- the likelihood is
\begin{equation}
\mathcal{L}(\theta) \propto \prod_{i} \exp \left( - \frac{(\gamma_{1,i}^\mathrm{d} - \gamma_{1,i}^\mathrm{m}(\theta))^2 + (\gamma_{2,i}^\mathrm{d} - \gamma_{2,i}^\mathrm{m}(\theta))^2}{2 \sigma_\gamma^2} \right),
\label{L_shear}
\end{equation}
where the product is over all pixels of our shear maps, superscripts $\mathrm{d}$ and $\mathrm{m}$ refer to the data and model respectively, and $\theta$ denotes the eNFW model parameters. \citet{2007ApJS..172..219L} found that each component of the intrinsic ellipticities of individual galaxies are Gaussian distributed, with a standard deviation of $\sigma_\mathrm{int} \sim 0.26$ across a wide range of sizes, magnitudes and redshifts. Thus, given a number of lensed galaxies, $N$, per pixel of a shear map, the contribution of intrinsic ellipticities to the average ellipticity of galaxies in that bin will be normally distributed with zero mean and standard deviation $\sigma_\gamma = 0.26/\sqrt{N}$. In our case, $N$ is set by the assumed number density of galaxies and the pixel size.

\begin{figure*}
        \centering
        \includegraphics[width=0.9\textwidth]{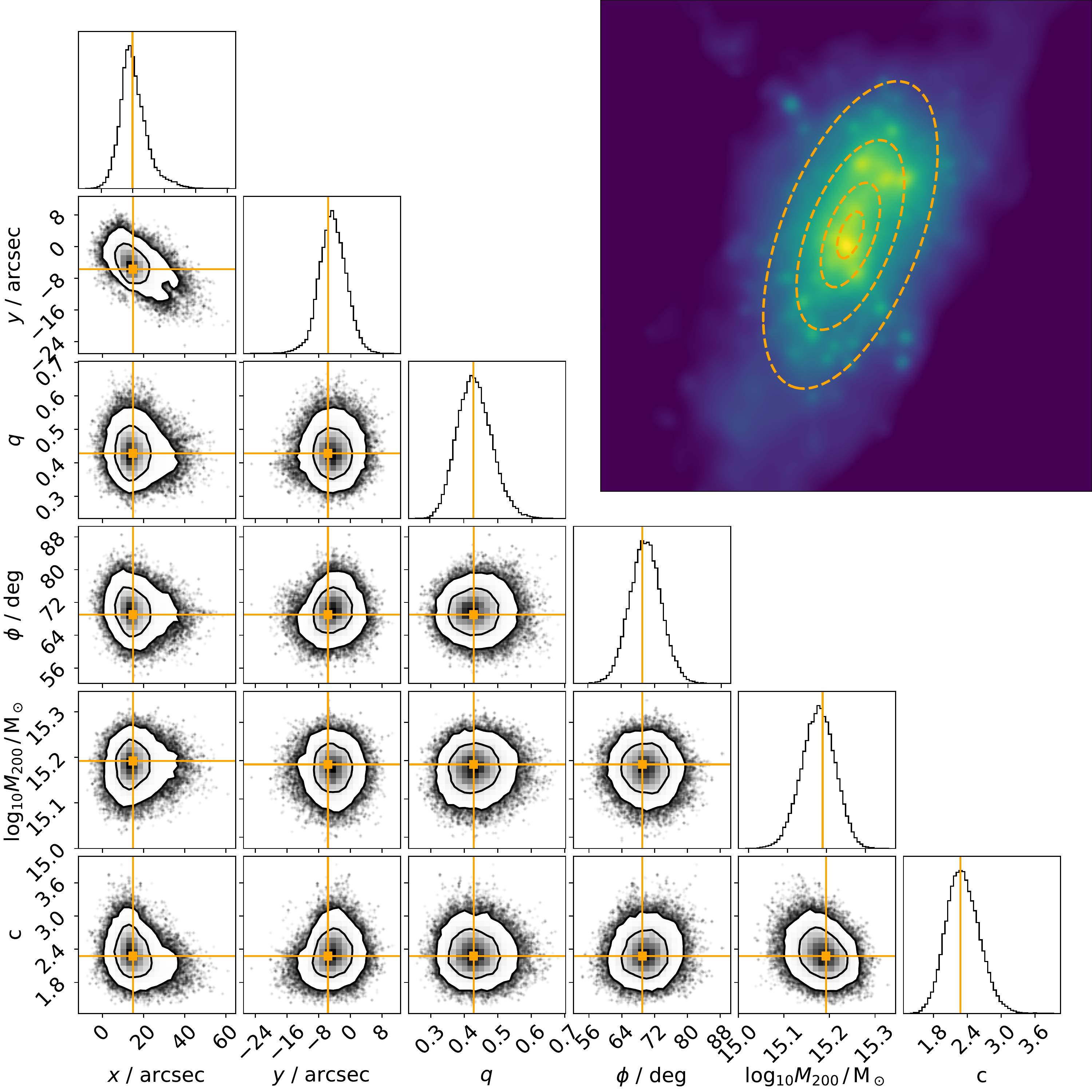}     
	\caption{An example corner plot, showing the posterior distribution for the eNFW parameters from fitting to noise-free shear data for one of our simulated clusters (this example is from the \bahamas-CDM simulation). The lens and source redshifts are $z_l = 0.375$ and $z_s = 2$, respectively, and the likelihood assumed that intrinsic galaxy shapes were the only source of noise, with 30 galaxies/arcmin$^2$. The orange lines denote the location of the best-fit (MAP) parameter values. In the top-right we show a map of the projected density field of this example cluster, with iso-convergence contours of the MAP eNFW model over-plotted as orange dashed lines. The side-length of the map in the top-right panel is 20 arcmin, slightly smaller than the 30 arcmin field over which the shear was fit.}
	\label{fig:example_eNFW_fit}
\end{figure*}

With our likelihood defined, we can now calculate a posterior distribution for the eNFW parameters, given some prior distribution for these parameters. The parameters in question are the $x$ and $y$ coordinates of the centre of the eNFW, the mass ($M_{200}$) and concentration ($c$) of the eNFW profile, and the axis ratio ($q$) and position angle ($\phi$ -- the angle from the $x$-axis to the major axis of the mass distribution). Due to the fact that $\phi$ is ill-defined when $q=1$, and to avoid complications associated with the periodicity of the parameter $\phi$, we actually parameterise the eNFW in terms of two components of ellipticity, related to $q$ and $\phi$ by
\begin{equation}
\epsilon_1 = \frac{1-q}{1+q} \cos 2 \phi, \\
\epsilon_2 = \frac{1-q}{1+q} \sin 2 \phi.
\label{eq:elliptical_components}
\end{equation}
We adopt Gaussian priors on the $x$ and $y$ coordinates of the halo centre, with a standard deviation of 5 arcmin, centred on the projected location of the most gravitationally bound particle within the galaxy cluster. We adopt a uniform prior on $\log M_{200}$, for $M_{200}$ values between $10^{12}$ and $10^{16} \msun$, and a uniform prior on $c$ in the range 1 to 20. Finally, we adopt uniform priors on each of $\epsilon_1$ and $\epsilon_2$ in the range -1 to 1, with the additional constraint that $\epsilon_1^2 + \epsilon_2^2 < 1$.

Using these priors, and the likelihood defined in Equation~\eqref{L_shear}, we use the package \textsc{emcee} \citep{ForemanMackey:2013io} to perform an MCMC fit to the shear maps from each of our simulated clusters. An example posterior distribution (for one of our CDM haloes) is shown as a corner plot in Fig.~\ref{fig:example_eNFW_fit}, with the maximum a posteriori (MAP) model parameters marked in orange. In the top-right of Fig.~\ref{fig:example_eNFW_fit} the convergence map for this cluster is plotted, with iso-convergence contours of the MAP model overlaid in orange. It is clear from this figure that a model with elliptical symmetry is simplistic when compared with the true projected mass distribution of this simulated cluster, owing to features like substructure in the lens that are not captured by the model. Nevertheless, the axis ratio and position angle of the model appear to have a strong correspondence with the projected density map. Also from this figure, we see that with SuperBIT-like data we can determine $q$ for an individual cluster to a precision of $\sigma_q = 0.05$ (the standard deviation of the marginalised $q$ posterior in this case). Note that the 100 most massive haloes from each of the CDM and SIDM1 \bahamas simulations span a range of halo masses of approximately $14.3 < \log_{10} M_{200} / \msun < 15.3$, and the standard deviation of the $q$ posterior distribution is found to increase with decreasing halo mass. For our sample of haloes, the relation between $\sigma_q$ and $M_{200}$ is well described by $\sigma_q = 0.10 - 0.21 \log_{10} (M_{200} / 10^{15} \msun)$.\footnote{This clearly cannot be valid for considerably higher masses, given that $\sigma_q$ would become negative at masses greater than approximately $10^{15.5} \msun$. While we only have 5 clusters with $M_{200} > 10^{15} \msun$, there is a slight indication that this relationship begins to plateau towards $\sigma_q \approx 0.05$ at the high mass end.}

SuperBIT's ability to measure the shapes of individual galaxy clusters with reasonable precision opens up new ways of analysing cluster shapes. In contrast, current ground based weak lensing data -- which has a lower number density of usable weak lensing sources -- typically requires cluster shear fields to be stacked in order to get a reasonable signal-to-noise ratio. Stacking has clear drawbacks for studies of cluster shapes, because intrinsically elliptical clusters will appear round when stacked, unless some method is employed to align the clusters before stacking. This can be done, for example by assuming that the cluster's mass distribution is well-aligned with the distribution of cluster member galaxies \citep[see][for example]{2018MNRAS.475.2421S}, but then the shape of the stacked cluster mass distribution depends on both the intrinsic shape distribution of clusters and the level of alignment between the DM and stars within clusters. Being able to measure shapes for individual clusters allows things like the misalignment between cluster DM haloes and cluster galaxy distributions to be measured.

Because we fit to noise-free data, the MAP point in the eNFW parameter space takes on a special significance. The MAP parameters are independent of $\sigma_\gamma$, and correspond to those that we would infer for a particular mass distribution in the limit of infinite signal-to-noise weak lensing data. For the rest of this paper we therefore focus only on the MAP estimates of the axis ratios from the eNFW fits to our mock shear fields.

\subsection{Distribution of weak lensing axis ratios}

\begin{figure}
        \centering
        \includegraphics[width=\columnwidth]{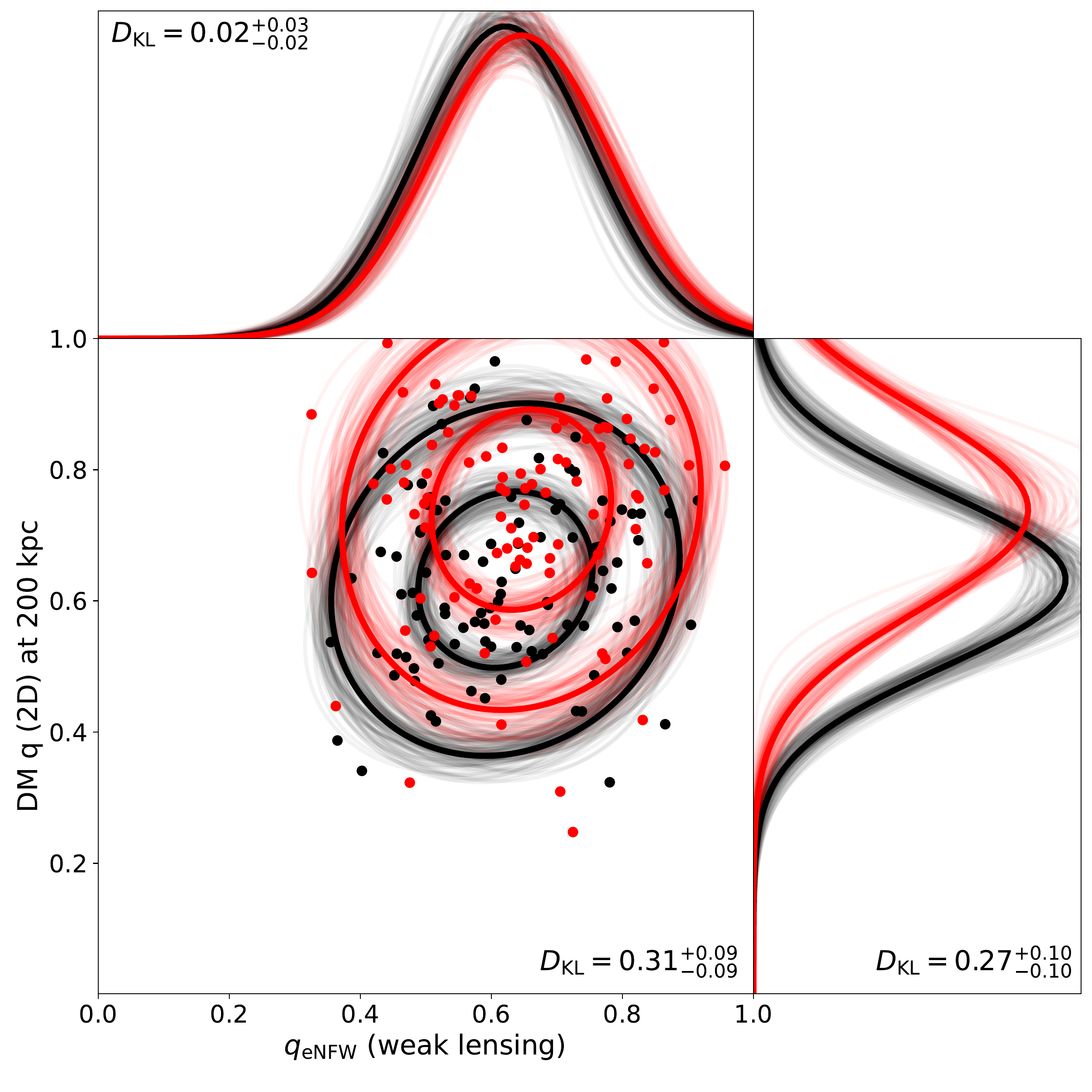}     
	\caption{The distributions of axis ratios measured by fitting eNFW profiles to shear fields, as well as from the 2D reduced inertia tensor for the DM at $R = 200 \kpc$, where the plot style is that same as in Fig.~\ref{fig:3D_joint_dist}. The joint distributions indicate that these two axis ratios are almost uncorrelated in both CDM and SIDM1.}
	\label{fig:WL_shape_dist}
\end{figure}

Having fit an eNFW model to the shear field for each of the 100 most massive \bahamas-CDM and \bahamas-SIDM1 clusters, we show the distribution of axis ratios in Fig.~\ref{fig:WL_shape_dist}. The CDM and SIDM1 distributions are almost indistinguishable, with a $D_{KL}$ value of just $0.02_{-0.02}^{+0.03}$, similar to that for the DM axis ratio measured in projection at $2 \mpc$. We also plot the 2D DM axis ratio measured at $200 \kpc$ -- which was already plotted in the left panel of Fig.~\ref{fig:2D_joint_dist} -- such that we can discuss the joint distribution of weak lensing shapes combined with 2D shapes at small radii in Section~\ref{sect:discussion_of_eInOut}.

\subsection{On what scale does weak lensing measure the halo shape?}
\label{sect:scale_of_WL}

Given that 2D reduced inertia tensor shapes for the DM are different between CDM and SIDM1 (i.e. Fig.~\ref{fig:2D_axis_ratios}) it is somewhat disappointing that the distributions of weak lensing shapes do not seem to reflect this difference. To understand why, it is useful to address the question ``on what radial scale does weak lensing measure the halo shape?''. Of course there need not be a unique answer to this that is valid for all haloes, because the eNFW fit attempts to describe the complete shear field (and so the complete mass distribution of the cluster) with an axis ratio that is independent of radius, while cluster mass distributions are more complex than this. The fact that the median axis ratios in Fig.~\ref{fig:2D_axis_ratios} vary slightly with radius (especially with SIDM) tells us that individual clusters cannot have axis ratios that are independent of radius, and Fig.~\ref{fig:2D_axis_ratios} overstates how smoothly the axis ratio varies with radius for individual systems, because it is a median of all systems.

\begin{figure}
        \centering
        \includegraphics[width=\columnwidth]{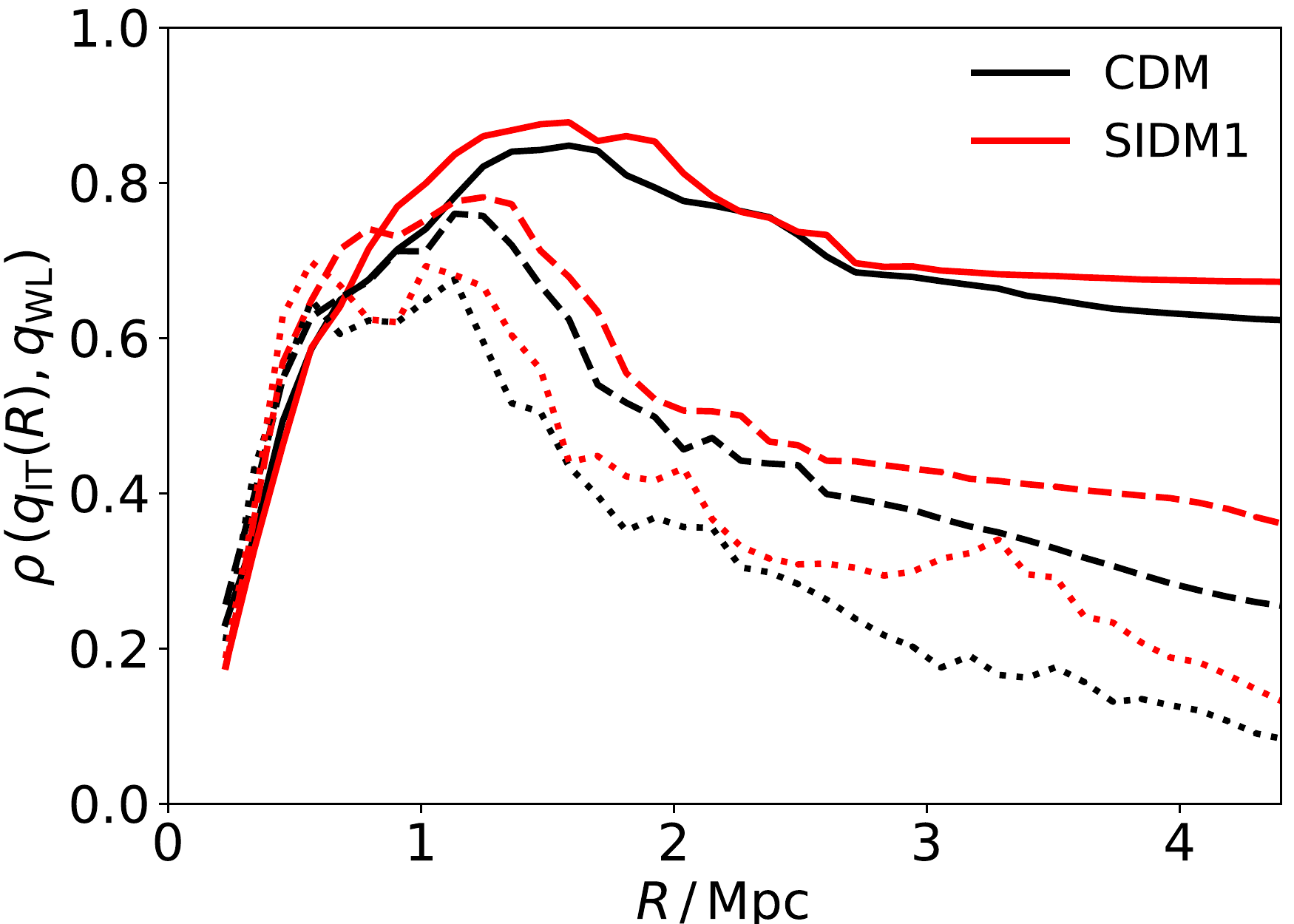}     
	\caption{The Pearson correlation coefficient between the weak lensing axis ratio and various definitions of 2D inertia tensor shapes (where the inertia tensor shape is evaluated at a radius of $R$). Solid lines are for the reduced inertia tensor, dashed lines for the standard inertia tensor, and dotted lines for the inertia tensor in shells. The weak lensing shapes are highly correlated with the reduced inertia tensor shapes, measured on a scale of $R \sim 1.5 \mpc$.}
	\label{fig:WL_vs_IT_correlation}
\end{figure}

To gain some insight into the radial scale on which weak lensing is most sensitive to the halo shape, we compare the axis ratios as a function of radius measured by the 2D inertia tensor, with the weak lensing axis ratios. For each DM model, and at each radius, we have 100 inertia tensor shapes, and the corresponding 100 weak lensing shapes. We calculate the Pearson correlation coefficient, $\rho$, between these inertia tensor shapes and weak lensing shapes, and plot them as a function of the radius at which the inertia tensor shape was calculated in Fig.~\ref{fig:WL_vs_IT_correlation}. The different line styles correspond to different inertia tensor definitions, and it is clear that the weak lensing shape is most correlated with the reduced inertia tensor shape measured at $R \sim 1.5 \mpc$, for which the correlation coefficient is $\rho \approx 0.85$. The fact that cluster weak lensing shapes are most closely related to an inertia tensor shape measured on the scale of $1.5 \mpc$ (approximately the virial radius of these $\lesssim 10^{15} \msun$ clusters at z=0.375), while SIDM changes the shape primarily of the inner halo (see Fig.~\ref{fig:3D_axis_ratios} and Fig.~\ref{fig:2D_axis_ratios}), explains why the weak lensing axis ratio distributions are so similar between CDM and SIDM1.

It is interesting to note that when measuring the inertia tensor in shells, it is shells at $R \sim 1 \mpc$ that are the most correlated with the weak lensing shape. This shift towards smaller radii when compared with the reduced inertia tensor is most likely driven by the fact that the contribution of different annuli to the total mass of an NFW cluster ($\propto R \times \Sigma(R)$) peaks at around the scale radius, which for clusters of this mass is around 500--800$\kpc$. The fact that the reduced inertia tensor outperforms the standard inertia tensor in terms of predicting the weak lensing shape, with a correlation coefficient approaching 1, suggests that the weighting given to different radii by the reduced inertia tensor, roughly matches the weighting of different radii to the overall weak lensing shear field.

We also note that the 2D reduced inertia tensor shapes at $1.5 \mpc$ and the weak lensing shapes are not just highly correlated, the axis ratios measured by the two methods are approximately the same. For example, with CDM: $\left< q_\mathrm{IT}(R=1.5\mpc) \right> = 0.62$ and $\left< q_\mathrm{WL} \right> = 0.63$, while the standard deviations are $\sigma_{q,\mathrm{IT}} = 0.16$ and $\sigma_{q,\mathrm{WL}} = 0.14$, with a correlation coefficient between the two shape measurements of 0.84. This result -- that weak lensing can accurately measure the shape of a halo as it would be measured by a simulator using the inertia tensor -- is in contrast with the findings of \citet[][hereafter, \citetalias{2021MNRAS.500.2627H}]{2021MNRAS.500.2627H}, who found that weak lensing inferred axis ratios were typically much rounder than those calculated from the inertia tensor. Specifically, for a definition of ellipticity $\epsilon = (1-q^2)/(1+q^2)$,\footnote{This is different from the definition of ellipticity implicit in equation~\eqref{eq:elliptical_components}, which is $\epsilon = (1-q)/(1+q)$} they found that $\left< \epsilon_\mathrm{WL} / \epsilon_\mathrm{IT} \right> = 0.23$.

While we do not know the cause of this difference, there was no particular explanation in \har for why weak lensing would infer such round shapes for clusters, and we suspect our result (that weak lensing can measure a shape that is closely related to that from the inertia tensor) is correct. Differences between our work and \har include a significantly smaller field-of-view in \har (only out to about 40\% of the virial radius), and the use of so-called pseudo-elliptical NFW (pNFW) mass distributions. A pNFW is similar to an eNFW, in that it starts off with a spherically symmetric NFW profile, however, whereas the eNFW deforms an NFW into something with elliptical iso-convergence contours, the pNFW has elliptical iso-potential contours instead \citep{2002A&A...390..821G, 2007NJPh....9..447J}. 

The use of pNFW profiles is driven by the deflection angles being easier to compute, but they can sometimes correspond to unusual mass distributions, which can have ``dumbbell-shaped'' iso-convergence contours, and even regions with negative convergence. That said, \citet{2012A&A...544A..83D} found that these problems were primarily for large ellipticities, and that for ellipticities typical of simulated halos, pNFW mass distributions can provide an accurate description of an eNFW mass distribution. This means that the use of pNFW profiles is, in and of itself, unlikely to be the cause of the low ellipticities in \har. However, there are various ellipticity definitions in the literature surrounding pNFW profiles, and a conversion must be made between the axis ratio of the iso-potential contours and the axis ratio of the corresponding matter distribution, such that the \har results could stem from some mismatch in ellipticity definitions when using the pNFW.


\subsection{Fitting NFW profiles with radially-varying ellipticity}

Given that shape differences are most apparent in the inner halo, but that weak lensing-inferred axis ratios are dominated by the shapes of the outer halo, one possibility for distinguishing between CDM and SIDM1 using halo shapes is to fit a model where the inner and outer halo have separate ellipticities. We implement such a model within \textsc{PyAutoLens} \citep{pyautolens}, making use of the fact that \textsc{PyAutoLens} can use a ``CSE decomposition'' \citep{2021PASP..133g4504O} to calculate the deflection angles for elliptical NFW profiles. The idea of the CSE decomposition is to first represent an NFW profile as the sum of many cored steep ellipsoid (CSE) profiles. In projection, a CSE profile has a constant surface density at small radii, which then turns over (at a radius that we call $R_\mathrm{core}$) to fall off as $\Sigma \propto 1/R^3$ at large radii, details are in  \citet{2021PASP..133g4504O}.

\citet{2021PASP..133g4504O} showed that 44 CSE profiles could represent an NFW density profile to a fractional accuracy of 1 part in $10^4$ over the full range of radial scales that we are interested in, and this particular CSE decomposition \citep[for which the coefficients can be found in Table 1 of][]{2021PASP..133g4504O} is the one that we use here. The advantage of using many CSEs instead of an NFW profile, comes when one wishes to make the projected mass distribution elliptical. This is because the deflection angle field due to an elliptical NFW profile requires computationally expensive numerical integrals to be performed, whereas the deflection angles due to each CSE are analytic. As such, using a CSE decomposition can speed up deflection angle calculations for eNFW profiles by over two orders of magnitude when compared with the standard approach.

While not the originally intended purpose of the CSE decomposition, another potential benefit is the flexibility to have more complex angular structure than a radially-independent axis ratio and position angle. Each CSE has an associated $R_\mathrm{core}$, and by varying the axis ratio and/or position angle of the CSEs in a systematic way with the core radius, one can achieve NFW-like profiles which have a radially dependent axis ratio and/or have `twists' in the iso-convergence contours. Here we try one very simple version of this, having separate axis ratios and position angles for the inner and outer halo. Specifically, we add two new parameters to the six of the eNFW profile: $\epsilon_1^\mathrm{in}$ and $\epsilon_2^\mathrm{in}$, which are the two components of the ellipticity of the inner halo, related to the axis ratio and position angle of the inner halo through eq.~\eqref{eq:elliptical_components}. We call this model mass distribution an eInOutNFW profile.

\begin{figure*}
        \centering
        \includegraphics[width=0.9\textwidth]{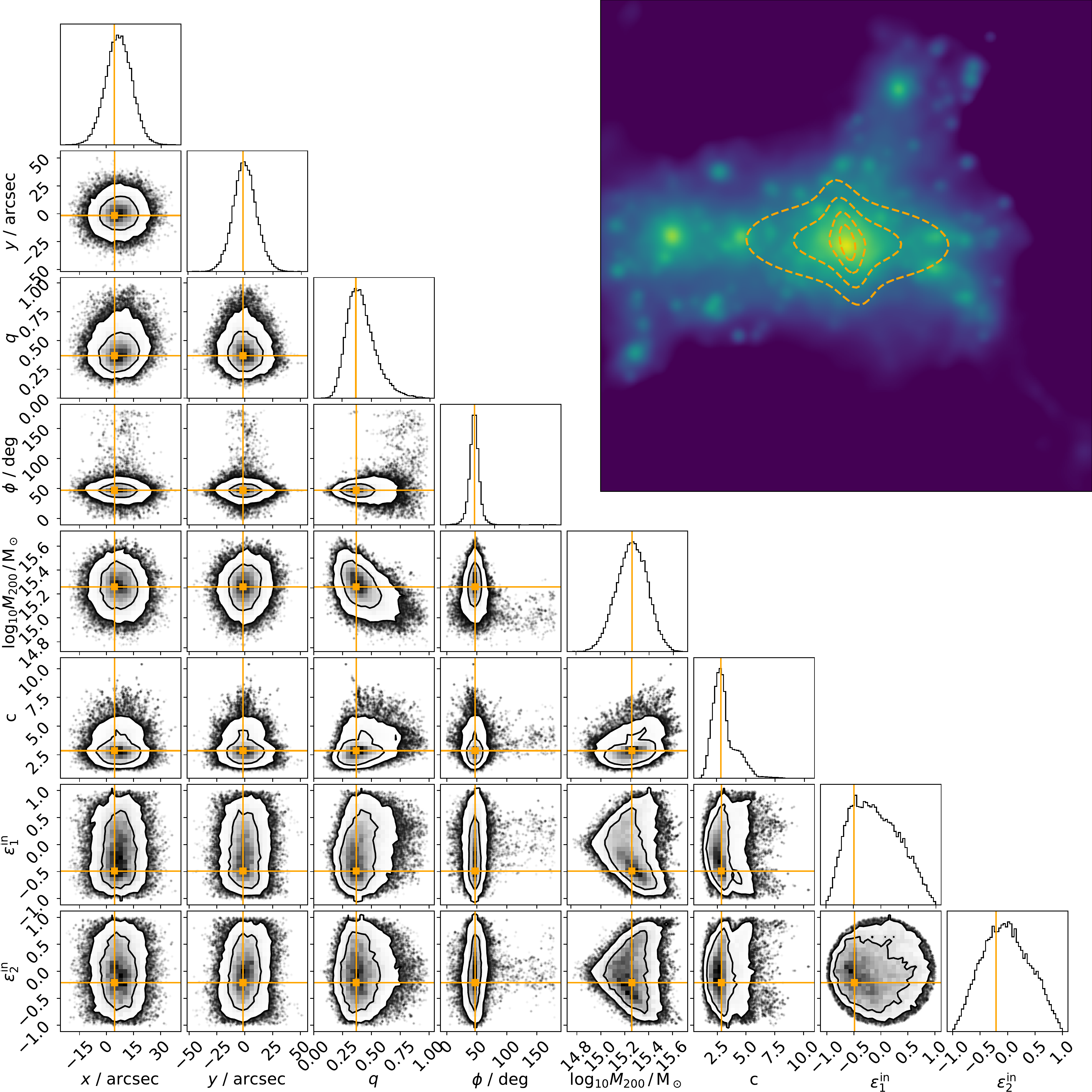}     
	\caption{An example corner plot, showing the posterior distribution for the eInOutNFW parameters from fitting to noise-free shear data for one of our simulated clusters (this example is a different halo from that in Fig.~\ref{fig:example_eNFW_fit}, though also from the \bahamas-CDM simulation). The lens and source redshifts are $z_l = 0.375$ and $z_s = 2$, respectively, and the likelihood assumed that intrinsic galaxy shapes were the only source of noise, with 30 galaxies/arcmin$^2$. The eInOutNFW profile used a fixed transition radius between the inner and outer halo of $R_q = 500 \kpc$. The orange lines denote the location of the best-fit (MAP) parameter values. In the top-right we show a map of the projected density field of this example cluster, with iso-convergence contours of the MAP eInOutNFW model over-plotted as orange dashed lines. The side-length of the map in the top-right panel is 20 arcmin ($7.5 \mpc$), slightly smaller than the 30 arcmin field over which the shear was fit.}
	\label{fig:example_qCutNFW_fit}
\end{figure*}

We show an example of fitting this model to a mock cluster shear field in Fig.~\ref{fig:example_qCutNFW_fit}, where we adopted a transition radius between the inner and outer halo of $R_q = 500 \kpc$. All CSEs with $R_\mathrm{core} < R_q$ had an ellipticity set by $\epsilon_1^\mathrm{in}$ and $\epsilon_2^\mathrm{in}$, while CSEs with larger $R_\mathrm{core}$ had an ellipticity given by  $\epsilon_1$ and $\epsilon_2$ (plotted as $q$ and $\phi$ in Fig.~\ref{fig:example_qCutNFW_fit}). Note that while one could allow $R_q$ to also be a free parameter in the fit, we found that degeneracies between parameters in the resulting posteriors made them hard to interpret.

There are a couple of noteworthy comments to make about Fig.~\ref{fig:example_qCutNFW_fit}. The first is that with SuperBIT-like data, the posterior distribution covers all possible inner ellipticities ($|\epsilon_1^\mathrm{in}|^2 + |\epsilon_2^\mathrm{in}|^2 < 1$). This means that such a model is not actually something that could be constrained with weak lensing data alone, and instead would require data more sensitive to the mass distribution on small scales, which could potentially come from strong lensing in the cluster core. The second observation we wish to make is that the eInOutNFW model does not produce a system with elliptical iso-convergence contours (i.e. the orange dashed lines in Fig.~\ref{fig:example_qCutNFW_fit}). This is because each CSE contributes at a range of radii, and so at any given radius the mass distribution comes from a sum of profiles with different axis ratios and position angles.

\begin{figure}
        \centering
        \includegraphics[width=\columnwidth]{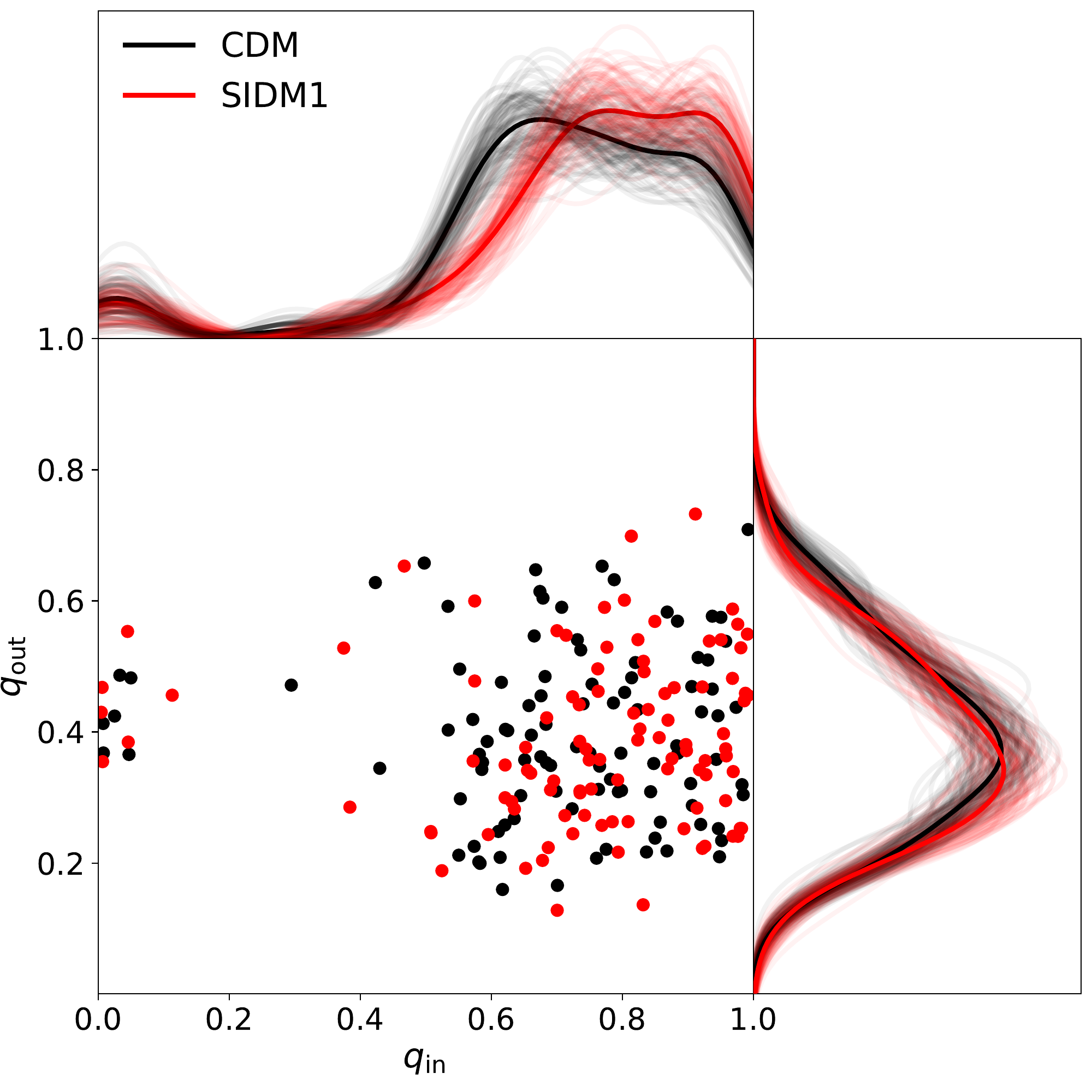}     
	\caption{The MAP parameters when fitting the eInOutNFW profile to shear data for our CDM and SIDM1 simulated clusters. The scatter plot shows the joint distribution of $q_\mathrm{in}$ and $q_\mathrm{out}$, while the other panels show the projections of this (i.e. the $q_\mathrm{in}$ distribution in the top panel, and the $q_\mathrm{out}$ distribution in the right panel). Unlike weak lensing and inertia tensor axis-ratios, the $q_\mathrm{in}$ and $q_\mathrm{out}$ distributions are not well approximated as Gaussians, so instead of plotting the best-fit Gaussians (as in, for example, Fig.~\ref{fig:WL_shape_dist}) we use kernel density estimation, with a Gaussian kernel with a standard deviation of 0.06 to plot the $q_\mathrm{in}$ and $q_\mathrm{out}$ distributions. We also show, as faded lines, the kernel density estimation of these distributions for 100 bootstrapped samples.}
	\label{fig:eInOutNFW_qDist}
\end{figure}

Even though the inner halo ellipticity parameters cannot be constrained from weak lensing data, it is nevertheless informative to look at the MAP parameters from our fits (because the mock shear field is noise free, and so the MAP parameters are a statement about the best-fit model to the shear field due to the intrinsic mass distribution). In Fig.~\ref{fig:eInOutNFW_qDist} we plot the distributions of inner and outer axis ratios, using $R_q = 500 \kpc$. The distributions of $q_\mathrm{in}$ and $q_\mathrm{out}$ are not well approximated as Gaussian, and so -- unlike in some previous figures -- we use kernel density estimation\footnote{We used \texttt{sklearn.neighbors.KernelDensity \citep{scikit-learn}}, with \texttt{kernel=`gaussian'} and \texttt{bandwidth=0.06}.} to visualise the distributions. We stress that the results for individual haloes were not very stable, in that they change quite considerable as $R_q$ is varied. However, it is encouraging that the behaviour seems to align with expectations, with distributions of outer axis ratios that are similar between CDM and SIDM1, but with an excess of SIDM1 systems that have $q_\mathrm{in} \gtrsim 0.7$. Note that in both CDM and SIDM1 there are systems with quite extreme inner axis ratios ($q_\mathrm{in} < 0.2$). Visual inspection of the convergence maps for these systems reveal them to be bimodal systems, with two distinct density peaks in the inner regions of the cluster.



\subsection{Discussion of using the eInOutNFW profile}
\label{sect:discussion_of_eInOut}

The eInOutNFW could be used as a cluster-scale mass model, with the ability to reveal information about the DM cross-section through the distribution of the inner halo axis ratios. The envisioned usage, would be for observational studies that try to fit to combined strong and weak lensing \citep[for example, using the hybrid-\textsc{Lenstool} software][]{2020MNRAS.493.3331N}. Such software assumes there is a single mass distribution that can explain both the observed strong and weak lensing, so if realistic mass distributions can have radially dependent ellipticities, it would be good to enable such mass distributions within the modelling.

That said, with the specific goal of constraining SIDM using cluster shapes, it is not clear that a single mass model to simultaneously explain both strong and weak lensing data is actually required. Returning our attention to Fig.~\ref{fig:WL_shape_dist}, we see that the shape of the projected mass distribution on small ($200 \kpc$) scales is almost completely uncorrelated with the weak lensing shape. As such, one does not gain much constraining power between CDM and SIDM1 by adding in weak lensing information, above what can be inferred from the 2D shape at $200 \kpc$ alone. If we take the 2D shape at $200 \kpc$ to be a proxy for what one would infer from fitting to strong lensing data, then inferring the shape of the inner halo by fitting to strong lensing data alone may be just as good as building a single model for the mass distribution that has a radially varying ellipticity, and fitting to combined strong and weak lensing data. With that somewhat pessimistic note about the role of weak lensing for this particular science case, we leave an investigation into using strong lensing cluster shapes as a probe of SIDM to future work.

\section{Conclusions}
\label{sect:conclusions}

We used a set of cosmological hydrodynamical simulations run with two different DM models -- CDM as well as SIDM with $1 \cmsg$ -- to investigate using the shapes of galaxy clusters as a test of the nature of DM. We used the reduced inertia tensor to measure the 3D minor-to-major axis ratios of the DM, gas and stars within clusters, finding that while DM shapes differed quite strongly (with SIDM leading to rounder DM distributions at the centres of haloes than CDM) the distribution of gas and star axis ratios were similar in the different DM scenarios. Unfortunately, this change in DM axis ratios is not reflected in the axis ratios of the projected mass distribution of clusters as they would be inferred from weak lensing data, primarily because the differences in 3D between CDM and SIDM are washed out somewhat in projection, and because weak lensing measures the halo shapes on roughly the scale of the virial radius, whereas the major differences in DM axis ratio happen in the inner halo. Our key findings can be summarised as:
%
\begin{itemize}
\item SIDM makes the DM in the inner regions of galaxy clusters more round, while leaving the shape of the star and gas distributions unchanged.
\item Despite the distribution of stellar axis ratios being the same in CDM and SIDM, information about the stellar shapes of galaxy clusters used in conjunction with the corresponding DM shapes make it easier to distinguish between these two DM models than DM shapes alone.
\item The differences between the distributions of DM shapes in CDM and SIDM are less pronounced in 2D projection than in 3D.
\item Balloon-borne imaging from the SuperBIT telescope should be able to measure the axis ratio of individual galaxy clusters, with a precision of approximately $\sigma_q = 0.1$ for clusters with a mass of $10^{15} \msun$, and a greater precision for more massive clusters.
\item The distribution of weak-lensing-inferred axis ratios is the same in CDM and SIDM. The primary driver of this is that cluster weak lensing measures the shape on a scale of $R \sim 1.5 \mpc$, whereas the shape differences are most apparent at $R \lesssim 500 \kpc$.
\end{itemize}
Having found that the simple procedure of fitting elliptical NFW profiles to the shear fields of clusters could not distinguish between CDM and SIDM, we introduced a more complex model, where the ellipticity of a cluster's mass distribution can be different at small and large radii. We found that when fitting this model to CDM and SIDM clusters, the distribution of inner axis ratios was shifted towards unity (meaning more round) in SIDM compared with CDM, as expected. However, weak lensing data alone would be insufficient to constrain the inner axis ratio with any meaningful precision, and so fitting this model would most likely require constraints from strong lensing data in order to distinguish between CDM and SIDM using the distribution of galaxy cluster shapes.

\section*{Data Availability}

The data underlying this article will be shared on reasonable request to the corresponding author.

\section*{Acknowledgments}

The authors thank Ellen Sirks, Kyle Oman, Richard Massey, Carlos Frenk, Spencer Everett, Jason Rhodes and David Harvey for helpful discussions. This research was carried out at the Jet Propulsion Laboratory, California Institute of Technology, under a contract with the National Aeronautics and Space Administration (80NM0018D0004). AR was supported in part by NASA grant 14-APRA14-0013 and acknowledges the support of the SuperBIT team. The High Performance Computing resources used in this investigation were provided by funding from the JPL Information and Technology Solutions Directorate.

The \bahamas-SIDM simulations used in this work were run on the DiRAC@Durham facility, managed by the Institute for Computational Cosmology on behalf of the STFC DiRAC HPC Facility (www.dirac.ac.uk). The equipment was funded by BEIS capital funding via STFC capital grants ST/K00042X/1, ST/P002293/1, ST/R002371/1 and ST/S002502/1, Durham University and STFC operations grant ST/R000832/1. DiRAC is part of the National e-Infrastructure.

This work made use of the following software packages: \href{https://github.com/astropy/astropy}{{Astropy}}
\citep{astropy1, astropy2},
\href{https://github.com/dfm/corner.py}{{Corner}}
\citep{ForemanMackey2016},
\href{https://emcee.readthedocs.io/en/stable/}{{emcee}}
\citep{ForemanMackey:2013io},
\href{https://github.com/matplotlib/matplotlib}{{Matplotlib}}
\citep{matplotlib},
\href{https://github.com/numpy/numpy}{{NumPy}}
\citep{numpy},
\href{https://github.com/Jammy2211/PyAutoLens}{{PyAutoLens}}
\citep{pyautolens},
and
\href{https://github.com/scipy/scipy}{{Scipy}}
\citep{scipy}.

\copyright 2022. California Institute of Technology. Government sponsorship acknowledged. 

\bibliographystyle{mnras}

\bibliography{bibliography}

\appendix

\section{Analytically projecting 3D shapes to 2D shapes}
\label{projecting_shapes}
Imagine that we have some 3D mass distribution, with inertia tensor $\mathbf{M}$ which has eigenvalues $a^2$, $b^2$ and $c^2$. Without loss of generality, let's assume that the associated eigenvectors are along the $x$, $y$ and $z$-axes, respectively (so the major axis is the $x$-axis and the minor axis is the $z$-axis). If we imagine that the mass distribution giving rise to the inertia tensor $\mathbf{M}$ is a multivariate Gaussian distribution of equal mass particles, then the density of particles (up to some overall normalisation) is entirely specified by $\mathbf{M}$.

To project this mass distribution along some line-of-sight, $\vect{e}_{z'}$, we need to find the positions of particles along two axes orthogonal to the $z'$ axis, which we label $x'$ and $y'$, with associated unit vectors $\vect{e}_{x'}$ and $\vect{e}_{y'}$. The variance of particle positions along the $x'$ axis will be $\sigma_{x'}^2 = \vect{e}_{x'}^\intercal \mathbf{M} \vect{e}_{x'}$, and that along the $y'$ axis will be $\sigma_{y'}^2\ = \vect{e}_{y'}^\intercal \mathbf{M} \vect{e}_{y'}$. Similarly the covariance between $x'$ and $y'$ coordinates of particles will be $ \sigma_{x'y'} = \vect{e}_{y'}^\intercal \mathbf{M} \vect{e}_{x'}$. The 2D inertia tensor of the projected particle distribution is simply
\begin{equation}
\mathbf{M}_{2\mathrm{D}} = 
\begin{pmatrix}
\sigma_{x'}^2 & \sigma_{x'y'} \\
\sigma_{x'y'}  & \sigma_{y'}^2
\end{pmatrix}.
\end{equation}
We can find the eigenvalues of $\mathbf{M}_{2\mathrm{D}}$, which we call $a_{2\mathrm{D}}$ and $b_{2\mathrm{D}}$. Then the 2D axis ratio is $q = b_{2\mathrm{D}} / a_{2\mathrm{D}}$.

\bsp
\label{lastpage}

\end{document}